\documentclass{vldb}
\usepackage{graphicx}
\usepackage{latexsym}
\usepackage{amsmath}
\usepackage{amssymb}
\usepackage{fancyvrb}
\usepackage{subfigure}
\usepackage{colortbl}
\usepackage{enumerate}
\usepackage{pifont}
\usepackage{stmaryrd}
\usepackage{textcomp}
\usepackage{multirow}
\graphicspath {{./} {Figures/}}

\begin{document}

\title{An Experimental Investigation of XML Compression Tools}

\numberofauthors{1}
\author{
\alignauthor
Sherif Sakr\\
       \affaddr{National ICT Australia }\\
       \affaddr{223 Anzac Parade, NSW 2052}\\
       \affaddr{Sydney, Australia}\\
       \email{Sherif.Sakr@nicta.com.au}
}

\maketitle

\begin{abstract}
This paper presents an extensive  experimental study of the
state-of-the-art of XML compression tools. The study reports the
behavior of nine XML compressors using a large  corpus of XML
documents which covers the different natures and scales of XML
documents. In addition to assessing and comparing the performance
characteristics of the evaluated XML compression tools, the study
tries to assess the effectiveness and practicality of using these
tools in the real world. Finally, we provide some guidelines and
recommendations which are useful for helping developers and users
for making an effective decision for selecting the most suitable
XML compression tool for their needs.
\end{abstract}

\section{Introduction}
The e\textbf{X}tensible \textbf{M}arkup \textbf{L}anguage
(\textbf{XML}) has been acknowledged to be one of the most useful
and important technologies that has emerged as a result of the
immensed popularity of HTML and the World Wide Web. Due to the
simplicity of its basic concepts and the theories behind, XML has
been used in solving numerous problems such as providing neutral
data representation between completely different architectures,
bridging the gap between software systems with minimal effort and
storing large volumes of semi-structured data. XML is often
referred as \emph{self-describing data} because it is designed in
a way that the schema is repeated for each record in the document.
On one  hand, this self-describing feature grants the XML great
flexibility and on the other  hand, it introduces the main
 problem of \emph{verbosity} of XML documents which results in huge document sizes.
This huge size lead to the fact that the amount of information
that has to be transmitted, processed, stored, and queried is
often larger than that of other data formats.  Since XML usage is
continuing to grow and large repositories of XML documents are
currently pervasive, a great demand for efficient XML compression
tools has been exist. To tackle this problem, several research
efforts have proposed the use of XML-conscious compressors which
exploits the well-known structure of XML documents to achieve
compression ratios that are better than those of general text
compressors. The usage of XML compressing tools has many
advantages such as: reducing the network bandwidth required for
data exchange, reducing the disk space required for storage and
minimizing the main memory requirements of processing and querying
XML documents.

Experimental evaluation and comparison of different techniques and
algorithms which deals with the same problem is a crucial aspect
especially in applied domains of computer science. This paper
presents an extensive experimental study for evaluating the
state-of-the-art of XML compression tools. We examine the
performance characteristics of nine publicly available XML
compression tools against a wide variety of data sets that
consists of 57 XML documents. The web page of this study
\cite{XMLCOMPBENCH} provides access to the test files, examined
XML compressors and the detailed results of this study.

The remainder of this paper is organized as follows. Section
\ref{SecCompressionTools} briefly introduces the XML compression
tools examined in our study and classifies them in different ways.
Section \ref{SecDataSet} presents the data sets used
 to perform the experiments.  Section \ref{SecTestingEnvironments} describes the test environments. Detailed and consolidated results of our experiments
are presented in Section \ref{SecExperiments} before we draw our
final conclusions in Section \ref{SecConclusion}.
\section{Survey of XML Compression Tools}
\label{SecCompressionTools}
\subsection{Features and Classifications}
A very large number of XML compressors have been proposed in the
literature of recent years. These XML compressors can be
classified with respect to three main characteristics. The first
classification is based on their awareness of the structure of the
XML documents. According to this classification, compressors are
divided into two main groups:
\begin{itemize}
    \item \textbf{General Text Compressors}: Since XML data are stored as text files, the first logical
approach for compressing XML documents was to use the traditional
general purpose text compression tools. This group of XML
compressors \cite{Gzip,Bzip,PPMPaper} is  \emph{XML-Blind}, treats
XML documents as usual plain text documents and
applies the
    traditional text compression techniques \cite{D1}.
    \item \textbf{XML Conscious Compressors}: This group of
    compressors are designed to take the advantage of the
    awareness of the XML document \emph{structure} to achieve better
    compression ratios over the general text compressors. This group of
    compressor can be further classified according to their
    dependence on the availability of the schema information of
    the XML documents as follows:\begin{itemize}
        \item \emph{Schema dependent compressors} where both of
        the encoder and decoder must have access to the document
        schema information to achieve the compression process \cite{DTDPPM,XAUST,MillauPaper,rngzip}.
        \item \emph{Schema independent compressors} where the
        availability of the schema information is not required to
        achieve the encoding and decoding processes \cite{XMill,XMLPPMPaper,SCMPPM,XWRT}.
    \end{itemize}
    Although schema dependent compressors may be, \emph{theoretically}, able  to  achieve
slightly higher compression ratios, they are not preferable or
commonly used in practice because there is no guarantee that the
schema
    information of the XML documents is always available.
\end{itemize}
The second classification of XML compressor is based on their
ability of supporting queries.
\begin{itemize}
    \item \textbf{Non-Queriable (Archival) XML Processor}: This group of the XML
    compressors does not allow any queries to be processed over the compressed
    format \cite{XMill,XMLPPMPaper,XAUST,Exalt,SCMPPM}. The main focus of this group is to achieve the
    highest compression ratio. By default, general purpose text compressors belong to the non-queriable
    group of compressors.
    \item \textbf{Queriable XML Processor}: This group of the XML
    compressors allow queries to be processed over their compressed
    formats \cite{XGrind,QXTPaper,XseqPaper}. The compression ratio of this group is usually worse than that of the archival XML compressors. However, the main focus of this group is to avoid full document
    decompression during query execution. In fact, the ability to perform direct queries on compressed XML formats is important for
many applications which are hosted on resource-limited computing
devices such as mobile devices and GPS systems. By default, all
queriable compressors are XML conscious compressors as well.
\end{itemize}
The third classification considers whether the compression schemes
operate in an online or offline manner.
\begin{itemize}
    \item \textbf{Online Compressors}: are able to stream the
    compressed data to the decoder i.e the decode is able to
    begin the process of decompression before the encode has
    finished transmitting the compressed data.
    \item \textbf{Offline Compressors}: don't allow the decoder to
    begin the decompression process until the entire compressed
    file has been received.
\end{itemize}
The \emph{online} feature of XML compression tools could be very
important for the scenarios where the users are heavily exchanging
compressed XML documents over networks. In these scenarios, online
decompression processors can effectively decrease the network
latency during the transmission process.

Table \ref{tblCompressorFeatures} lists the symbols that indicate
the features of each XML compressor included in the list of Table
\ref{tblCompressorList}.
\begin{table}
\centering
\begin{tabular}{|l|l|}
  \hline
  Symbol & Description  \\
  \hline
  G & General Text Compressor \\
  S & Specific XML Compressor \\
  D & Schema dependent Compressor \\
  I & Schema Independent Compressor \\
  A & Archival XML Compressor \\
  Q & Queriable XML Compressor \\
  O & Online XML Compressor \\
  F & Offline XML Compressor \\
  \hline
\end{tabular}
\caption{\footnotesize{Symbols list of XML compressors features}}
\label{tblCompressorFeatures}
\end{table}

\subsection{Examined Compressors}
In our study we considered, to the best of our knowledge,
\emph{all} XML compression tools which are fulfilling the
following conditions:
\begin{enumerate}
    \item Is publicly and freely  available either in the form of open source
codes or binary versions. \item Is a schema-independent. As
previously mentioned, the set of compressors which is not
fulfilling this condition is not commonly used in practice . \item
Be able to run under our Linux version of operating system.
\end{enumerate}

Table \ref{tblCompressorList} lists the surveyed XML compressors
and their features where the \textbf{Bold} font is used to
indicate the compressors which are fulfilling our conditions and
included in our experimental investigation. The border line
between the upper section and the lower section of  Table
\ref{tblCompressorList} is used to differentiate between the
non-queriable (upper section) and queriable (lower section) sets
of the XML compressors. The three compressors (\emph{DTDPPM,
XAUST, rngzip}) have not been included in our study because they
do not satisfy Condition 2. Although its source code is available
and it can be successfully compiled, XGrid did not satisfy
Condition 3. It always gives a fixed run-time error message during
the execution process. The rest of the list (11 compressors) don't
satisfy Condition 1. The status of a lack of source code/binaries
for a large number of the XML compressors proposed in literature,
to the best of our search efforts and contact with the authors,
and especially from the queriable class
\cite{XseqPaper,XCQPaper,QXTPaper} was a bit disappointing for us.
This has limited  a subset of our initially planned experiments
especially those which targeted towards assessing the performance
of evaluating the results of XML queries over the compressed
representations. In the following we give a brief description of
each examined compressor. \vspace{0.05in}
\newline\textbf{General Text Compressors}
 numerous algorithms have
been devised over the past decades to efficiently compress text
data. In our evaluation study we selected three compressors which
 are considered to be the best representative implementations of the most popular and efficient text compression techniques. We
selected gzip \cite{Gzip}, bzip2 \cite{Bzip} and PPM
\cite{PPMPaper} compressors to represent this group.
\vspace{0.05in}
\newline\textbf{XMill}
in \cite{XMill} Liefke and Suciu have presented the first
implementation of an XML conscious compressor. In XMill, both of
the structural and data value parts of the source XML document are
collected and compressed separately . In the structure part, XML
tags and attributes are encoded in a dictionary-based fashion
before passing it to a back-end general text compression scheme.
Data values are grouped into homogenous and semantically related
containers
 according to their path and data type. Each
container is then compressed separately using specialized
compressor that is ideal for the data type of this container. In
the latest versions of the XMill source distribution, the
intermediate binaries of the compressed format can be passed to
one of three alternative back-end general purpose compressor:
gzip, bzip2 and PPM. In our experiments we evaluated the
performance of the three alternative back-ends independently.
Hence, in the rest of the paper we refer to the three alternative
back-ends with the names \emph{XMillGzip}, \emph{XMillBzip} and
\emph{XMillPPM} respectively. \vspace{0.05in}
\newline\textbf{XMLPPM} is considered as an adaptation of the general purpose \emph{P}rediction by \emph{P}artial \emph{M}atching compression scheme (PPM) \cite{PPMPaper}. In \cite{XMLPPMPaper}, Cheney has presented XMLPPM as a streaming XML compressor which uses a
\emph{M}ultiplexed \emph{H}ierarchical PPM \emph{M}odel called
(MHM). The main idea of this MHM model is to use four different
PPM models for compressing the different XML symbols: element,
attribute, character and miscellaneous data. \vspace{0.05in}
\newline\textbf{SCMPPM}
is described by Adiego et al. in  \cite{SCMPaper} as a variant of
the XMLPPM compressor. It combines a technique called
\emph{S}tructure \emph{C}ontext \emph{M}odelling (SCM) with the
PPM compression scheme. It uses a bigger set of PPM models than
XMLPPM as it uses a separate model to compress the text content
under each element symbol. \vspace{0.05in}
\newline\textbf{XWRT} is presented by Skibinski et al. in \cite{XWRTPaper}. It applies a
dictionary-based compression technique called \emph{X}ML
\emph{W}ord \emph{R}eplacing \emph{T}ransform. The idea of this
technique is to replace the frequently appearing  words with
references to the dictionary which is obtained by a preliminary
pass over the data. XWRT submits the encoded results of the
preprocessing step to three alternative general purpose
compression schemes: gzip, LZMA and PPM. \vspace{0.05in}
\newline\textbf{Axechop} is presented by Leighton et al. in \cite{AXECHOPPaper}. It divides the
source XML document into structural and data segments. The MPM
compression algorithm is used to generate a context-free grammar
for the structural segment which is then passed to an adaptive
arithmetic coder. The data segment contents are organized into a
series of containers (one container for each element) before
applying the \emph{B}urrows-\emph{W}heeler \emph{T}ransformation
(BWT) compression \cite{BWT} over each container.
\newline\textbf{Exalt} in \cite{EXALTThesis}, Toman has presented an idea of applying
a syntactical-oriented approach for compressing XML documents. It
is similar to AXECHOP in that it utilized the fact that XML
document could be represented using a context-free grammar. It
uses the grammar-based codes encoding technique introduced by
Kieffer and Yang in \cite{KYGrammar} to encode the generated
context-free grammars. \vspace{0.05in}
\begin{table}
\centering \scriptsize
\begin{tabular}{|l|c|c|}
  \hline
  Compressor & Features & Code Available\\
  \hline
  \textbf{GZIP (1.3.12)} \cite{Gzip} &  GAIF & Y   \\
  \textbf{BZIP2 (1.0.4)} \cite{Bzip} & GAIF  & Y   \\
  \textbf{PPM (j.1)} \cite{PPM} & GAIF  & Y   \\
  \textbf{XMill (0.7)} \cite{XMillc} &  SAIF & Y   \\
  \textbf{XMLPPM (0.98.3)} \cite{XMLPPM}&  SAIO  & Y   \\
  \textbf{SCMPPM (0.93.3)}\cite{SCMPPM}&  SAI  & Y   \\
  \textbf{XWRT (3.2)} \cite{XWRT}& SAI  & Y   \\
  \textbf{Exalt (0.1.0)}\cite{Exalt}&  SAIF  & Y   \\
  \textbf{AXECHOP}\cite{AXECHOPPaper}&  SAIF  & Y   \\
  DTDPPM \cite{DTDPPM}&  SADO  & Y   \\
  XAUST\cite{XAUST}&  SAD  & Y   \\
  rngzip \cite{rngzip}& SQD  & Y   \\
  Millau\cite{MillauPaper}&  SADO  & N   \\
  XComp \cite{XCOMP}&  SAIF  & N   \\
  \hline
  XGrind \cite{XGrind}& SQIO  & Y   \\
  XBzip  \cite{XBZIPPaper} &  SQI  & N \\
  XQueC \cite{XQueC}& SQI &   N    \\
  XCQ \cite{XCQPaper}& SQIO &   N    \\
  XPress \cite{XPRESSPaper}&  SQIO  & N   \\
  XQzip \cite{XQZIPPaper} &  SQI  & N   \\
  XSeq \cite{XseqPaper} &  SQI  & N   \\
  QXT \cite{QXTPaper} &  SQI  & N   \\
  ISX \cite{ISXPaper} & SQI &  N   \\
  \hline
\end{tabular}
\caption{\footnotesize{XML Compressors List}}
\label{tblCompressorList}
\end{table}

\section{Our Corpus}
\label{SecDataSet}
\subsection{Corpus Characteristics}  \label{SecDataCharacteristics} Determining the XML
files that should be used for evaluating the set of XML
compression tools is not a simple task. To provide an extensive
set of experiments for assessing and evaluating the performance
characteristics of the XML compression tools, we have collected
and constructed a large corpus of XML documents. This corpus
contains a wide variety of XML data sources and document sizes.
Table \ref{tblDataSet} describes the characteristics of our
corpus. \emph{Size} denotes the disk space of XML file in MBytes.
\emph{Tags} represents the number of distinct tag names in each
XML document. \emph{Nodes} represents the total number of nodes in
each XML data set. \emph{Depth} is the length of the longest path
in the data set. \emph{Data Ratio} represents the percentage of
the size of data values with respect to the document size in each
XML file. The documents are selected to cover a wide range of
sizes where the smallest document is 0.5 MB and the biggest
document is 1.3 GB. The documents of our corpus can be classified
into four categories depending on their characteristics:
\begin{itemize}
    \item \textbf{Structural documents}
 this group of documents has no data contents at
    all. 100 \% of each document size is preserved to its
    structure information. This category of documents is used to assess the claim of XML
conscious compressors on using the well known structure of XML
documents for achieving higher compression ratios on the
structural parts of XML documents. Initially, our corpus consisted
of 30 XML documents. Three of these documents were generated by
using our own implemented Java-based random XML generator. This
generator produces completely random XML documents to a
parameterized arbitrary depth with only \emph{structural}
information (no data values). In addition, we created a
\emph{structural} copy for each document of the other 27
\emph{original} documents - with data values - of the corpus.
Thus, each \emph{structural} copy captures the structure
information of the associated XML \emph{original} copy and removes
all data values. In the rest of this paper, we refer to the
documents which include the data values as \emph{original}
documents and refer to the documents with no data values as
\emph{structural} documents. As a result, the final status of our
corpus consisted of 57 documents, 27 \emph{original} documents and
30 \emph{structural} documents. The size of our own 3 randomly
generated documents (R1,R2,R3) are indicated in Table
\ref{tblDataSet} and the size of the \emph{structural} copy of
each \emph{original} version of the document can be computed using
the following equation:
\begin{center}
$size(structural) = (1 - DR) * size(Original)$
\end{center}
where $DR$ represents the data ratio of the document.

    \item \textbf{Textual documents}: this category of documents
    consists of simple structure and high ratio of its contents is preserved to the data values. The ratio of the data
    contents of these documents represent more than 70\% of the  document size.
    \item \textbf{Regular Documents} consists mainly of regular
    document structure and short data contents. This document category reflects the
    XML view of relational data. The data ratio of these documents is in the
    range of
between
    40 and 60 percent.
    \item \textbf{Irregular documents} consists of documents that
    have very deep, complex and irregular structure. Similar to
    purely structured documents, this document category is
    mainly focusing on evaluating the efficiency of compressing irregular structural information of XML documents.
\end{itemize}
\subsection{Data Sets}
Our data set consists of the following documents: \vspace{0.05in}
\newline\textbf{EXI-Group} is a variant collection of XML documents included in the testing framework
of the Efficient XML Interchange Working Group \cite{EXIDataSet}.
\vspace{0.05in}
\newline\textbf{XMark-Group} the XMark documents model an auction database with deeply-nested
elements. The XML document instances of the XMark benchmark are
produced by the \emph{xmlgen}  tool of the XML benchmark project
\cite{XMarkDataSet}. For our experiments, we generated three XML
documents using three increasing scaling factors. \vspace{0.05in}
\newline\textbf{XBench-Group}
presents a family of benchmarks that captures different XML
application characteristics \cite{XBenchDataSet}. The databases it
generates come with two main models: 1) Data-centric (DC) model
contains data that are not originally stored in XML format such as
e-commerce catalog data and transactional data 2) Text-centric
(TC) model which represents text data that are more likely stored
as XML. Each of these two models can be represented either in the
form of a single document (SD) or multiple documents (MD). In
short, these two levels of classifications are combined to
generate four database instances: TCSD, DCSD, TCMD, DCMD. In
addition, XBench can generate databases with 4 different sizes:
small (11MB), normal (108MB) and large (1GB) and huge (10GB). In
our experiments, we only use TCSD and DCSD instances of the small
and normal sizes. \vspace{0.05in}
\newline\textbf{Wikipedia-Group} Wikipedia offers free copies of all content to interested users \cite{WikipediaDataSet}.  For our corpus, we
selected five samples of  the XML dumps with different sizes and
characteristics. \vspace{0.05in}
\newline\textbf{DBLP} presents the famous database of bibliographic information of computer science
journals and conference proceedings.  \vspace{0.05in}
\newline\textbf{U.S House} is a legislative document which provides information about the ongoing work of the U.S. House of
Representatives. \vspace{0.05in}
\newline\textbf{SwissProt} is a protein sequence database which describes
the DNA sequences. It provides a high level of annotations and a
minimal level of redundancy. \vspace{0.05in}
\newline\textbf{NASA} is an astronomical database which is constructed  by converting legacy flat-file formats into XML documents and then making them available to the
public. \vspace{0.05in}
\newline\textbf{Shakespeare} represents the gathering of a collection of marked-up Shakespeare
plays into a single XML file. It contains many long textual
passages. \vspace{0.05in}
\newline\textbf{Lineitem} is an XML representation of the transactional
relational database benchmark (TPC-H). \vspace{0.05in}
\newline\textbf{Mondial} provides the basic statistical information
on countries of the world. \vspace{0.05in}
\newline\textbf{BaseBall} provides the complete baseball statistics
of all players of each team that participated in the 1998 Major
League. \vspace{0.05in}
\newline\textbf{Treebank} is a large
collection of parsed English sentences from the Wall Street
Journal. It has a very deep, non-regular and recursive structure.
\vspace{0.05in}
\newline\textbf{Random-Group} this group of documents has been
generated using our own implementation of a Java-based random XML
generator. This generator is designed in a way to produce
\emph{structural} documents with very random, irregular and deep
structures according to its input parameters for the number of
unique tag names, maximum tree level and document size. We used
this XML generator for producing three documents with different
size and characteristics. The main aim of this group is to
challenge the examined compressors and assess the efficiency of
compressing the structural parts of XML documents.
\begin{table*}
\centering \scriptsize
\begin{tabular}{|l|l|r|r|r|r|r|}
  \hline
  Data Set Name & Document Name & Size (MB) & Tags  & Number of Nodes & Depth & Data Ratio\\
  \hline
  \multirow {5}{*}{EXI \cite{EXIDataSet}} & EXI-Telecomp.xml &  0.65&39  & 651398 & 7 &0.48\\
   &EXI-Weblog.xml&  2.60& 12 & 178419 & 3 &0.31\\
   &EXI-Invoice.xml& 0.93 &52  &78377&7&0.57\\
   &EXI-Array.xml& 22.18 & 47 & 1168115 &10&0.68\\
   &EXI-Factbook.xml& 4.12 & 199 & 104117 & 5 &0.53\\
   &EXI-Geographic Coordinates.xml& 16.20 & 17 & 55 & 3 &1\\
  \hline
  \multirow {3}{*}{XMark \cite{XMarkDataSet}} & XMark1.xml & 11.40 & 74 & 520546 & 12 &0.74\\
   &XMark2.xml & 113.80 & 74 & 5167121 & 12 &0.74\\
   &XMark3.xml &  571.75 & 74 & 25900899 & 12 &0.74\\
  \hline
  \multirow {4}{*}{XBench \cite{XBenchDataSet}} & DCSD-Small.xml&  10.60 & 50 & 6190628 & 8 &0.45\\
   &DCSD-Normal.xml& 105.60 & 50 & 6190628 & 8 &0.45\\
   &TCSD-Small.xml&  10.95 & 24 & 831393 & 8 &0.78\\
   &TCSD-Normal.xml&  106.25 & 24 & 8085816 & 8 &0.78\\
  \hline
  \multirow {5}{*}{Wikipedia \cite{WikipediaDataSet}} & EnWikiNews.xml & 71.09 & 20 & 2013778 & 5 &0.91\\
   &EnWikiQuote.xml& 127.25 &20  & 2672870 & 5 &0.97\\
   &EnWikiSource.xml& 1036.66 & 20 & 13423014 & 5 &0.98\\
   &EnWikiVersity.xml& 83.35 & 20 & 3333622 & 5 &0.91\\
   &EnWikTionary.xml& 570.00 & 20 & 28656178 & 5 &0.77\\
  \hline
  DBLP   & DBLP.xml&  130.72 & 32  & 4718588 & 5 &0.58\\
  \hline
  U.S House  &USHouse.xml & 0.52 & 43 & 16963 & 16& 0.77 \\
  \hline
  SwissProt & SwissProt.xml& 112.13  & 85 &13917441  & 5& 0.60 \\
  \hline
  NASA  &NASA.xml & 24.45 & 61 & 2278447 & 8&  0.66\\
  \hline
  Shakespeare  & Shakespeare.xml & 7.47 &22  & 574156 & 7&  0.64\\
  \hline
  Lineitem  & Lineitem.xml & 31.48 & 18 &2045953  & 3&  0.19\\
  \hline
  Mondial  &Mondial.xml & 1.75 & 23 & 147207 & 5 &  0.77\\
  \hline
  BaseBall & BaseBall.xml & 0.65  & 46 & 57812 & 6 &  0.11\\
  \hline
  Treebank  & Treebank.xml& 84.06 & 250  & 10795711 & 36 &0.70\\
  \hline
  \multirow {3}{*}{Random} & Random-R1.xml& 14.20 & 100  & 1249997 & 28 &0\\
   &Random-R2.xml& 53.90 & 200  & 3750002  & 34 &0\\
   &Random-R3.xml& 97.85 & 300 & 7500017 & 30 & 0\\
  \hline
\end{tabular}
\caption{\footnotesize{Characteristics of XML data sets }}
\label{tblDataSet}
\end{table*}
\section{Testing Environments}
\label{SecTestingEnvironments} To ensure the consistency of the
performance behaviors of the evaluated XML compressors, we ran our
experiments on two different environments. One environment with
high computing resources and the other with considerably limited
computing resources. Table \ref{tblSDPE} lists the setup details
of our high resources environment and Table \ref{tblSDLE} lists
the setup details of the limited one.
\begin{table}[h]
\centering
\begin{tabular}{|l|l|}
  \hline
  \textbf{Operating System} & Ubuntu 7.10 (Linux 2.6.22 Kernel) \\
  \hline
  \textbf{CPU} & Intel Core 2 Duo E6850 CPU  \\ & 3.00 GHz, FSB 1333MHz
\\& 4MB L2 Cache\\
  \hline
  \textbf{Hard Disk} &  Seagate ST3250820AS\\
  & 250 GB\\
  \hline
  \textbf{RAM} &  4 GB\\
  \hline
  \textbf{Compilers} &  gcc/g++ 4.1\\
  \hline
\end{tabular}
\caption{\footnotesize{Setup details of the powerful resources
environment}} \label{tblSDPE}
\end{table}

\begin{table}[h]
\centering
\begin{tabular}{|l|l|}
  \hline
  \textbf{Operating System} & Ubuntu 7.10 (Linux 2.6.20 Kernel) \\
  \hline
  \textbf{CPU} &  Intel Pentium 4\\ &2.66GHz,  FSB 533MHz\\ &512KB L2 Cache\\
  \hline
  \textbf{Hard Disk} &  Western Digital WD400BB \\
  & 40 GB\\
  \hline
  \textbf{RAM} &  512 MB\\
  \hline
  \textbf{Compilers} &  gcc/g++ 4.1\\
  \hline
\end{tabular}
\caption{\footnotesize{Setup details of the low resources
environment}} \label{tblSDLE}
\end{table}
\newpage
\section{Experiments}
\label{SecExperiments} We evaluated the performance
characteristics of XML compressors by running them through an
extensive set of experiments. The setup of  our experimental
framework was very challenging and complex. The details of this
experimental framework is described as follows:
\begin{itemize}
    \item We evaluated 11  XML compressors:
3 general purpose text compressors (gzip, bzip2, PPM) and 8 XML
conscious compressors (XMillGzip, XMillBzip, XMillPPM XMLPPM,
SCMPPM, XWRT, Exalt, AXECHOP). For our main set of experiments, we
evaluated the compressors under their default settings. The
rational behind this is that the default settings are considered
to be the recommended settings from the developers of each
compressors and thus can be assumed as the best behaviour. In
addition to this main set of experiments, we run additional set of
experiments with \emph{tuned parameters} for the highest value of
the level of compression parameter provided by some compressors
(gzip, bzip2, PPM, XMillPPM, XWRT). That means in total we run 16
\emph{variant} compressors. The experiments of the tuned version
of XWRT could be only be performed on the high resource setup
because they require at least 1 GB RAM. \item Our corpus consists
of 57 documents: 27 \emph{original documents}, 27 \emph{structural
copies} and 3 randomly generated \emph{structural documents} (see
Section \ref{SecDataCharacteristics}).  \item We run the
experiments on two different platforms. One with limited computing
resources and the other with high computing resources. \item For
each combination of an XML test document and an XML compressor, we
run two different operations (\emph{compression - decompression}).
\item To ensure accuracy, all reported numbers for our time
metrics (\emph{compression time - decompression time}) (see
Section \ref{PerfMetrics})    are the average of five executions
with the highest and the lowest values removed.
\end{itemize}
The above details lead to the conclusion that our number of runs
was equal to 9120 on each experimental platform (16 * 57
* 2 * 5), i.e 18240 runs in total. \newline In addition to running this
huge set of experiments, we needed to find the best way to
collect, analyze and present this huge amount of experimental
results. To tackle this challenge, we created our own mix of Unix
shell and Perl scripts to run and collect the results of these
huge number of runs. In this paper, we present an important part
from results of our experiments. For full detailed results, we
refer the reader to the web page of this experimental study
\cite{XMLCOMPBENCH}.

\subsection{Errors}
\begin{figure}
  \includegraphics[width=0.45\textwidth,height=2in]{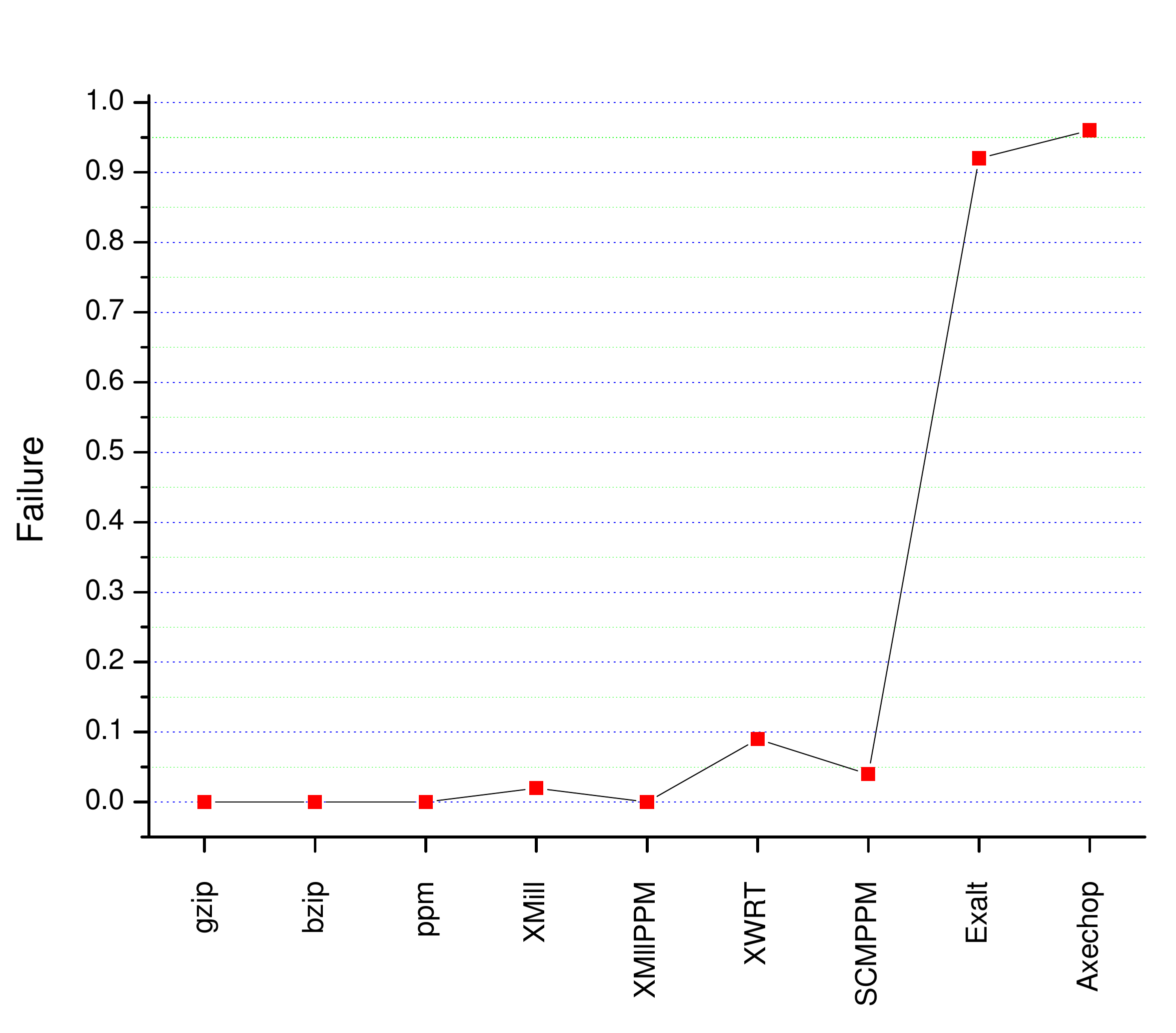}\\
  \caption{Percentage of unsuccessful runs of each compressor.}\label{FIG-Errors}
\end{figure}

During the run of our experiments, some tools failed to either
compress or decompress some of the documents in our corpus. We
consider the run as unsuccessful  if the compressor fails  to
achieve either of the encoding and decoding processes of the test
document. Thus, we had 57 runs for each compressor (one run per
document). Figure \ref{FIG-Errors} presents the percentage of
unsuccessful runs of each compressor. For a detailed list of the
errors generated during our experiments we refer to the web page
of this study \cite{XMLCOMPBENCH}. We have two main remarks about
the results of Figure \ref{FIG-Errors}:
\begin{itemize}
    \item The general purpose text compressors have shown complete stability. They were able to successfully perform the complete set of runs. They are
\emph{XML-Blind} thus require no knowledge of the
document-structure. Hence, they can deal with any XML document
even if it suffers from any syntax or well-formedness problems.
However, XML conscious compressors are very sensitive to such
problems. For example, some compressors which uses the Expat XML
parser such as XMLPPM will fail to compress any XML document which
uses external entity references if it does not have a \emph{dummy}
DTD declaration because the XML parser strictly applies the W3C
specification and will consider this document as not well-formed.
\item Except the latest version of XMLPPM (0.98.3), none of the
XML conscious compressors was able to execute the whole set of
runs successfully. Moreover, AXECHOP and Exalt compressors have
shown very poor \emph{stability}. They failed to run successful
decoding parts of many runs. They were thus excluded from any
consolidated results. Although an earlier version of XMLPPM
(0.98.2) suffered from some problem in decompressing the Wikipedia
data sets, the latest version of XMLPPM (0.98.3) released by
Cheney during the time of doing the experiments of this work has
fixed all earlier bugs and has shown to be the best XML conscious
compressor from the \emph{stability} point of view.
\end{itemize}
\subsection{Performance Metrics}
\label{PerfMetrics} We measure and compare the performance of the
XML compression tools using the following metrics:

\textbf{Compression Ratio}: represents the ratio between the sizes
of compressed and uncompressed XML documents.
\begin{center}
\small
Compression Ratio = (Compressed Size) / (Uncompressed Size)
\end{center}

\textbf{Compression Time}:  represents the elapsed time during the
compression process i.e the period of time between the start of
program execution on a document until all the data are written to
disk.

\textbf{Decompression Time}:  represents the elapsed time during
the decompression process i.e the period of time between the start
of program execution on reading the decompressed format of the XML
document until delivering the original document. \newline For all
metrics: \emph{the lower the metric value, the better the
compressor}.

\subsection{Experimental results}
\begin{figure}
  \includegraphics[width=0.45\textwidth,height=2in]{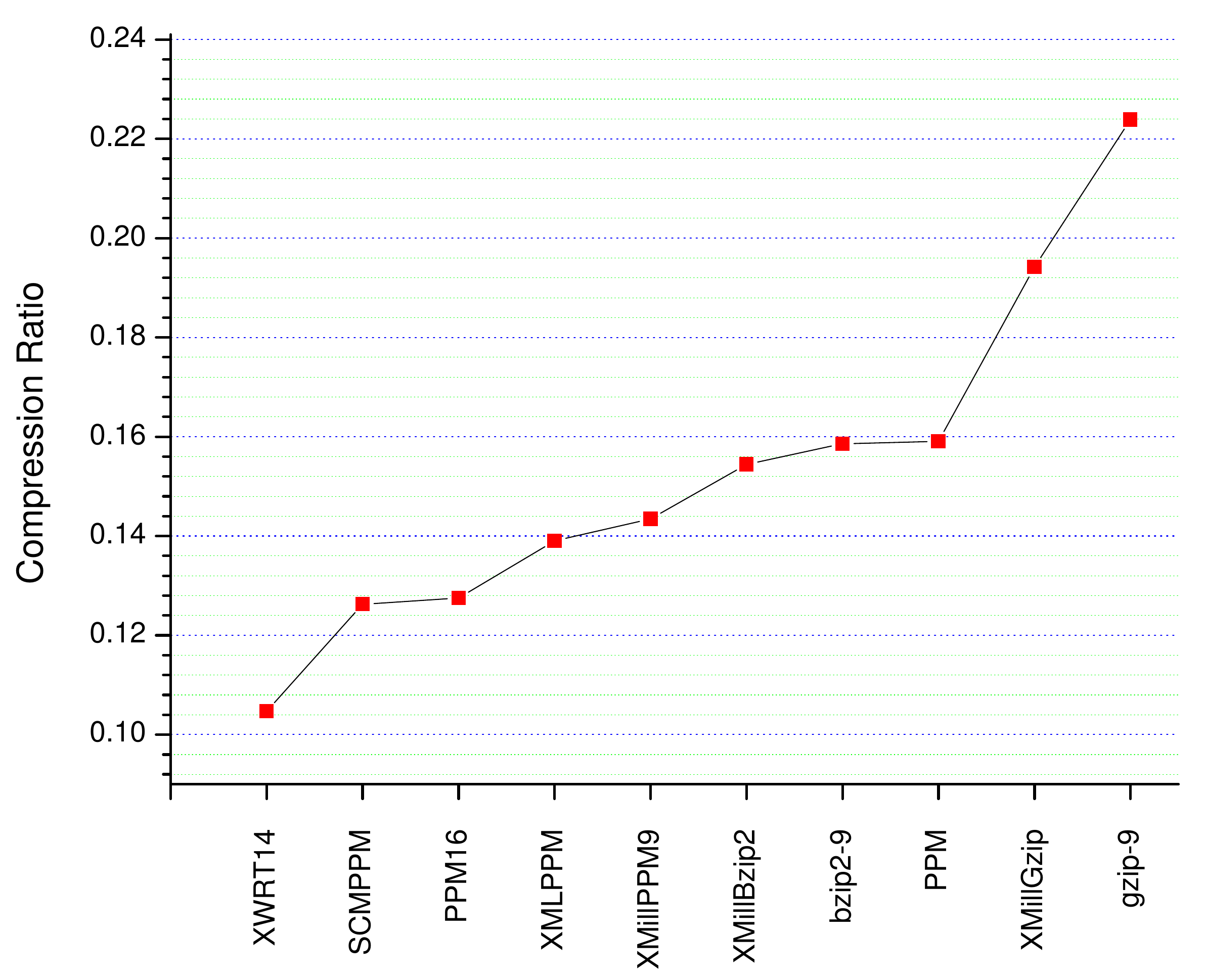}\\
  \caption{Average Compression Ratios (Tuned Parameters)}\label{FIG-P-O-CR-S}
\end{figure}

\begin{figure}
  \includegraphics[width=0.45\textwidth,height=2.1in]{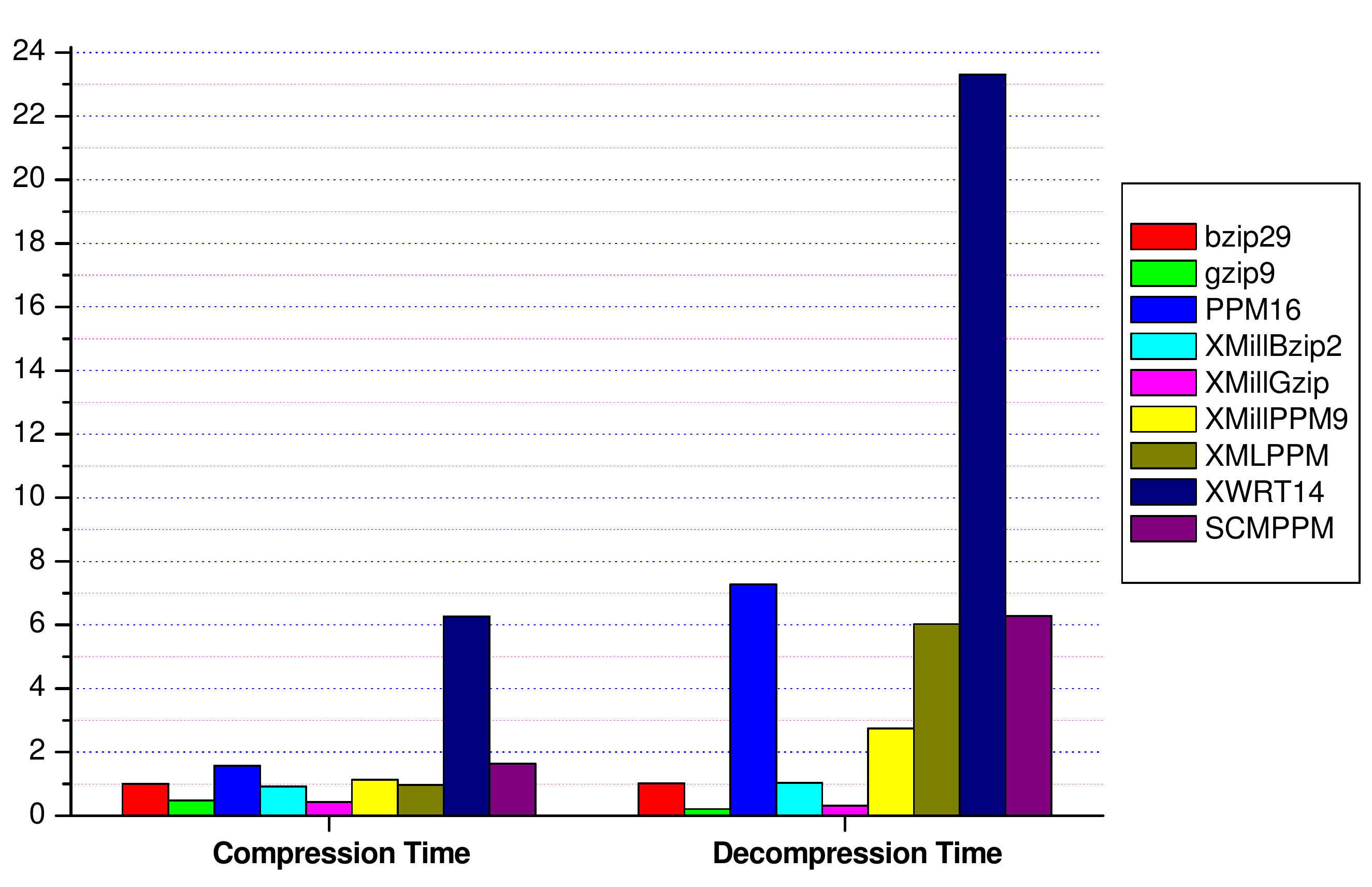}\\
  \caption{Average compression and decompression times over
\emph{original}  documents on high resources setup (Tuned
Parameters)}\label{FIG-P-S-O-T-D}
\end{figure}

\begin{figure*}
  \includegraphics[width=1\textwidth,height=3.5in]{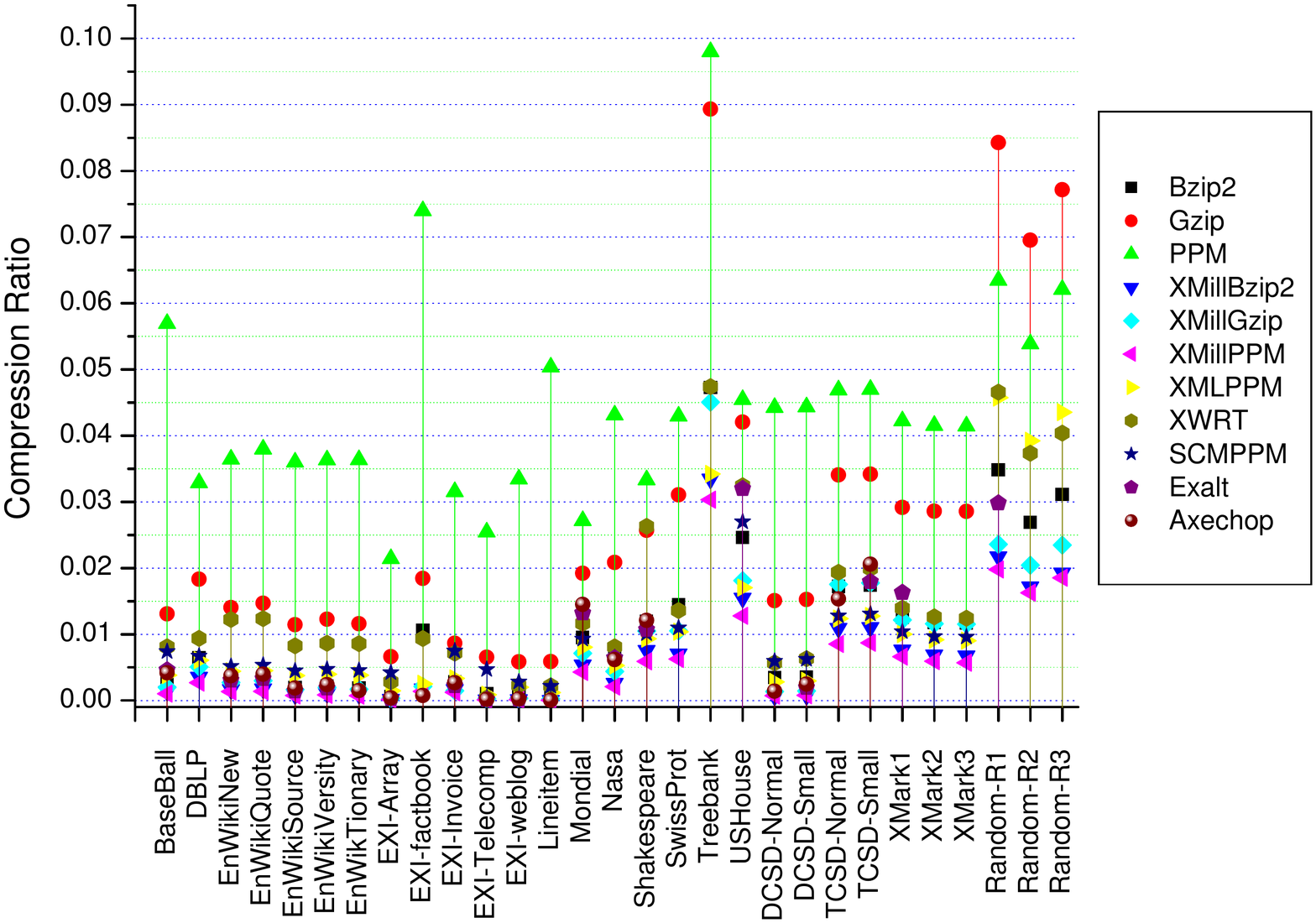}\\
  \caption{Detailed compression ratios of \emph{structural} documents}\label{FIG-S-S-CR-D}
\end{figure*}

\begin{figure*}
  \includegraphics[width=1\textwidth,height=3.5in]{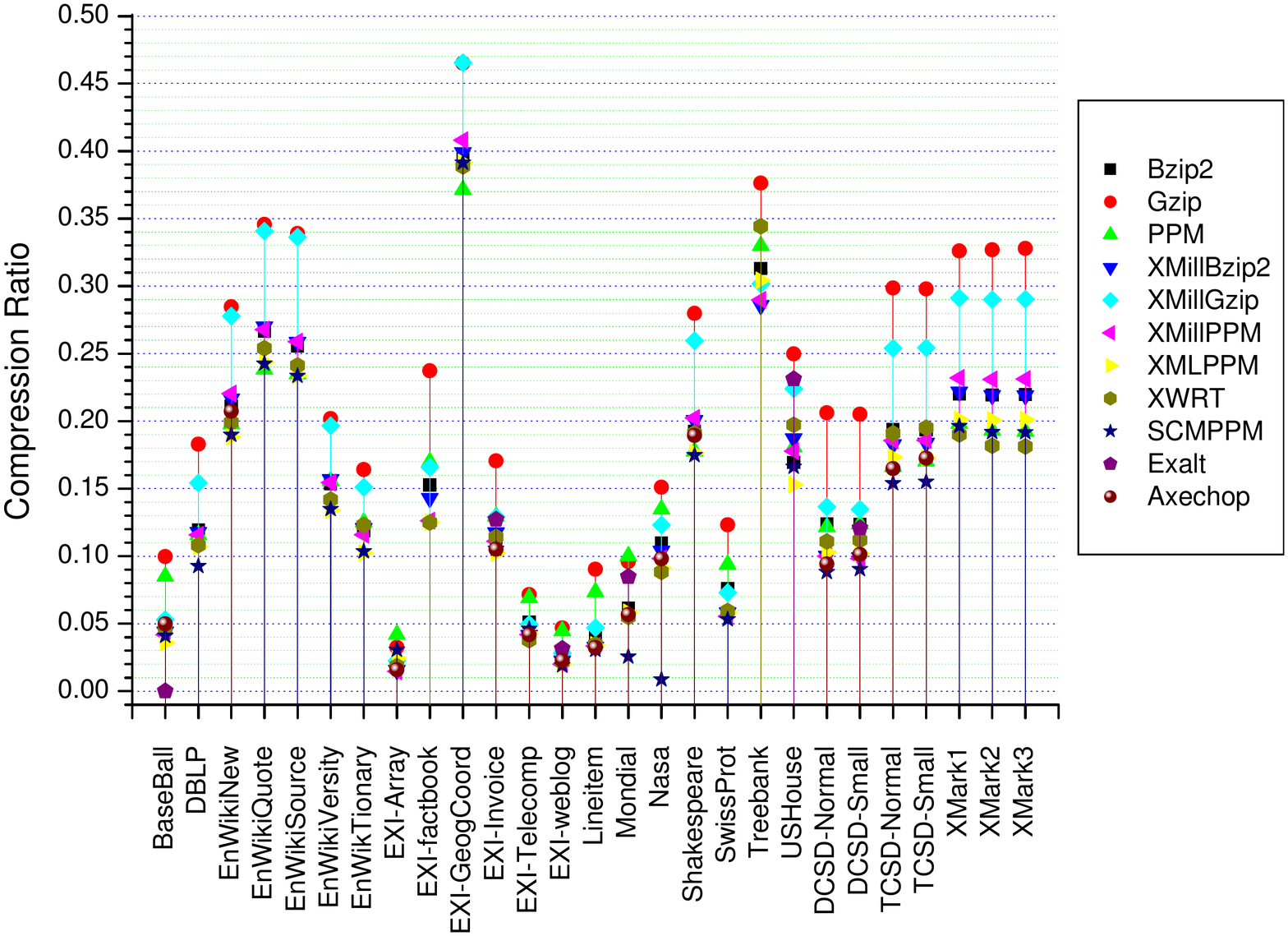}\\
  \caption{Detailed compression ratios of \emph{original} documents}\label{FIG-S-O-CR-D}
\end{figure*}

\begin{figure*}
\begin{minipage} [b]{0.62\textwidth}
\centering
\includegraphics[width=1\textwidth,height=3.3in]{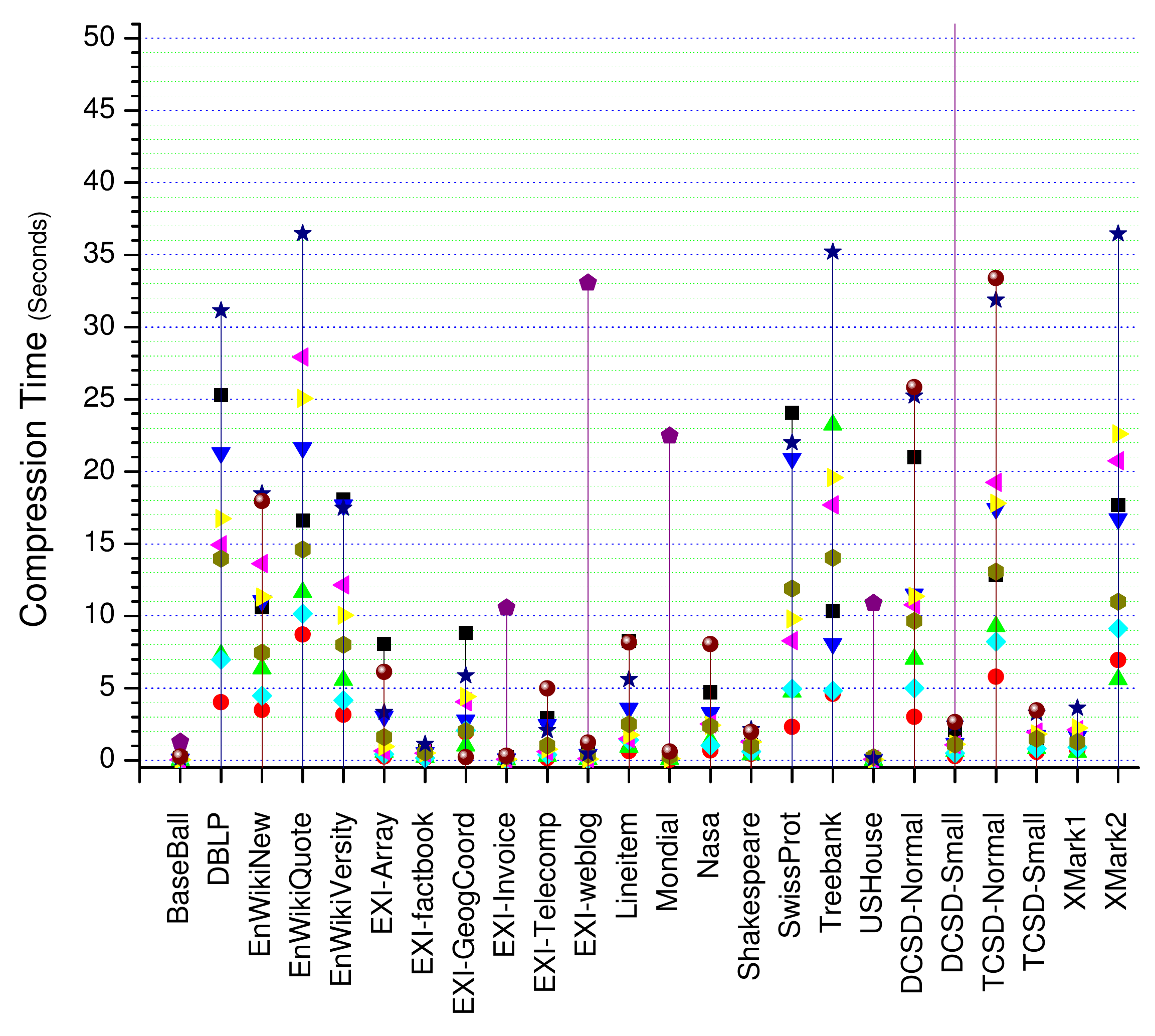}
\end{minipage}
\begin{minipage}[b]{0.35\textwidth}
\centering
\includegraphics[width=1\textwidth,height=3.3in]{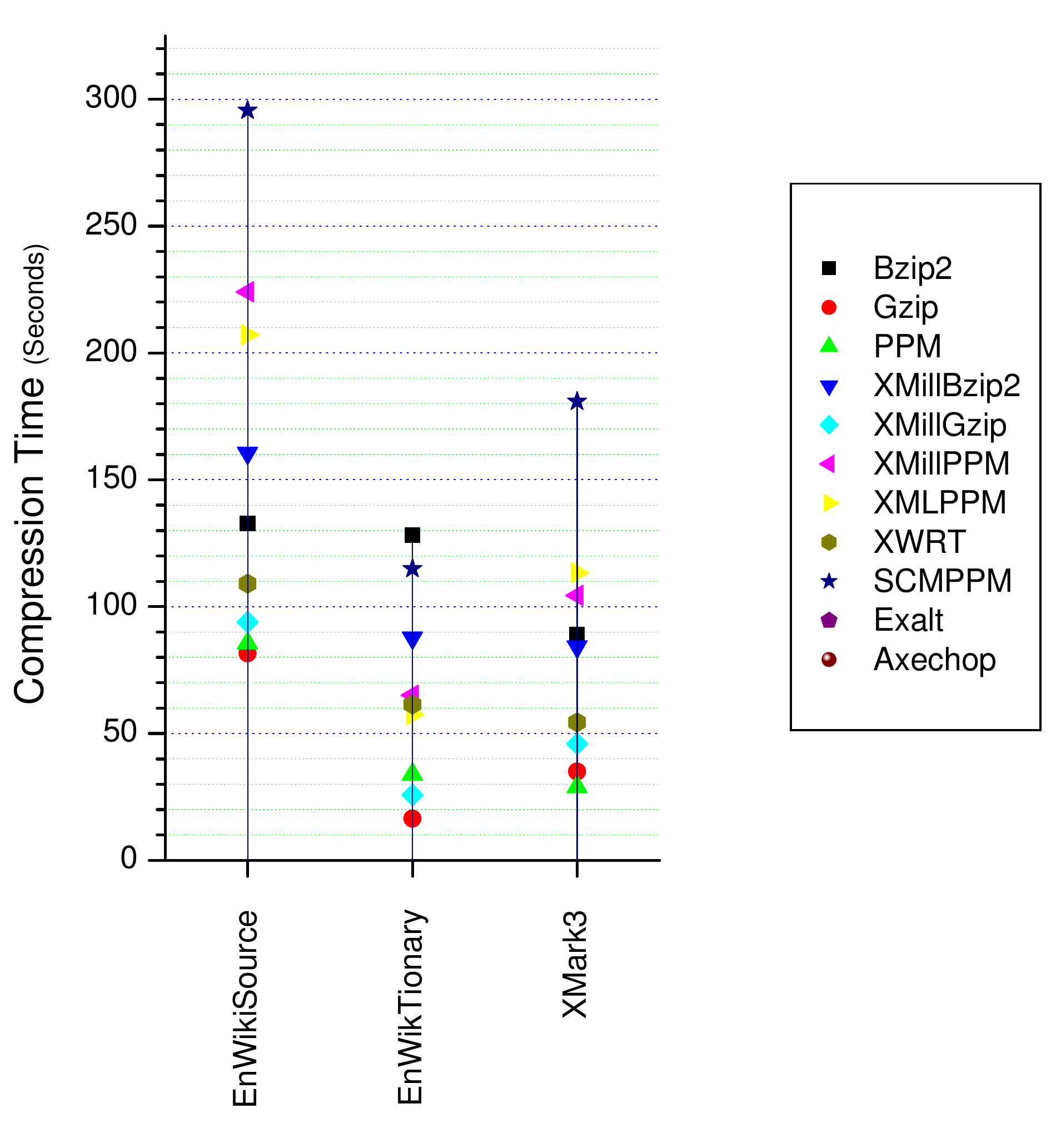}
\end{minipage}
\caption{Detailed compression times  on the high resources setup.}
\label{FIG-S-O-CT-D}
\end{figure*}

\begin{figure*}
\begin{minipage} [b]{0.62\textwidth}
\centering
\includegraphics[width=1\textwidth,height=3.3in]{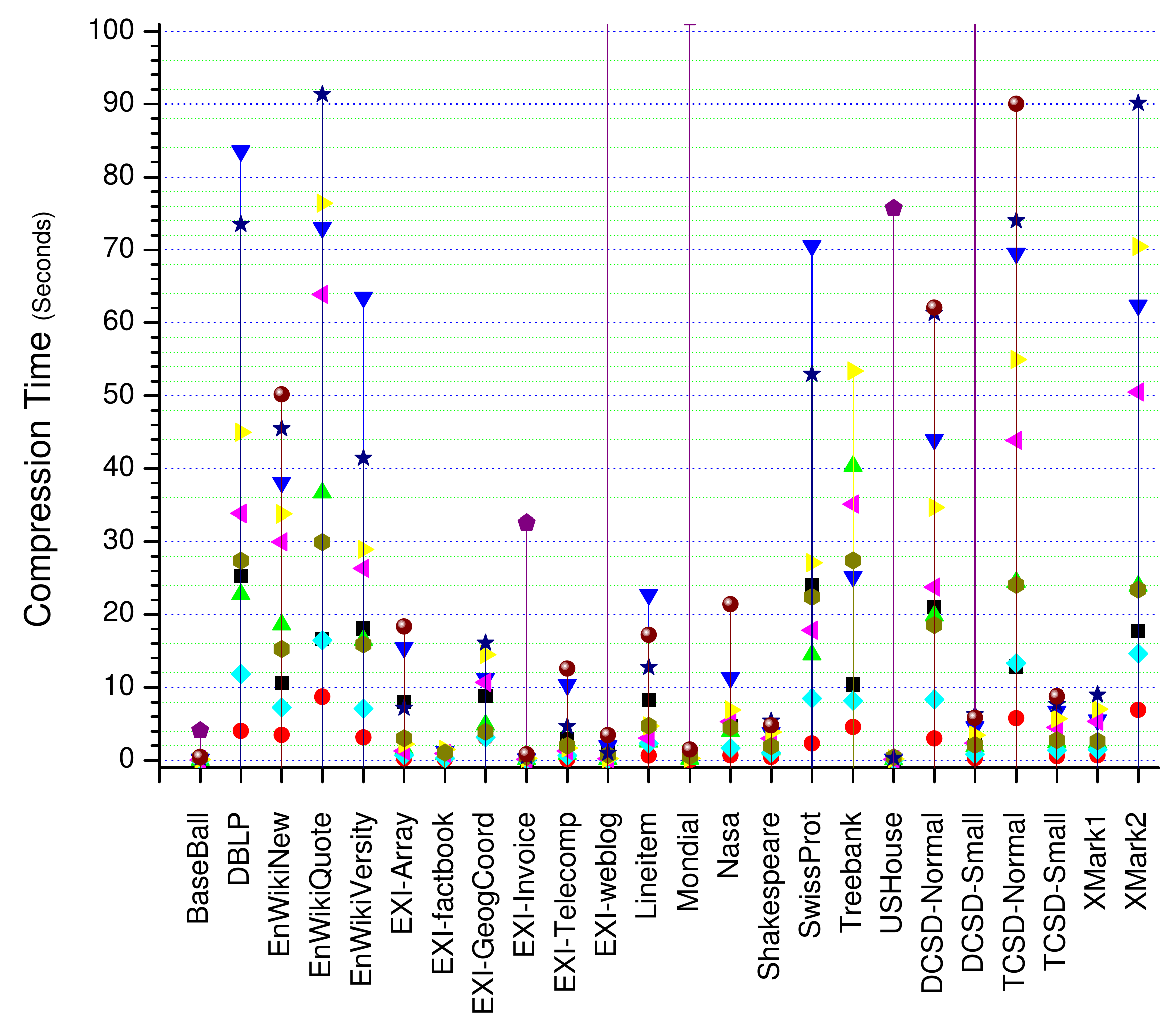}
\end{minipage}
\begin{minipage}[b]{0.35\textwidth}
\centering
\includegraphics[width=1\textwidth,height=3.3in]{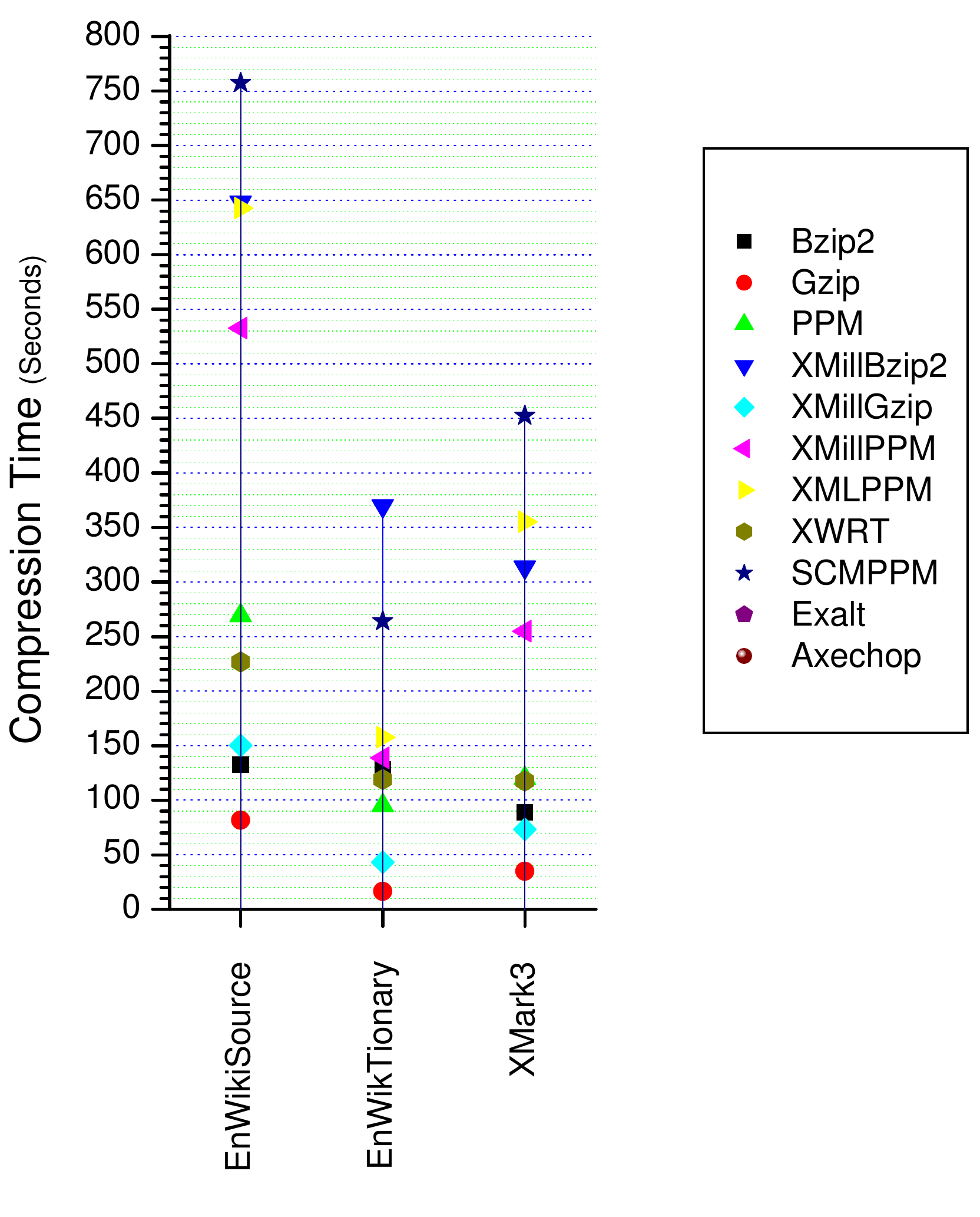}
\end{minipage}
\caption{Detailed compression times  on the limited resources
setup.} \label{FIG-C-O-CT-D}
\end{figure*}

\begin{figure*}
\begin{minipage} [b]{0.62\textwidth}
\centering
\includegraphics[width=1\textwidth,height=3.3in]{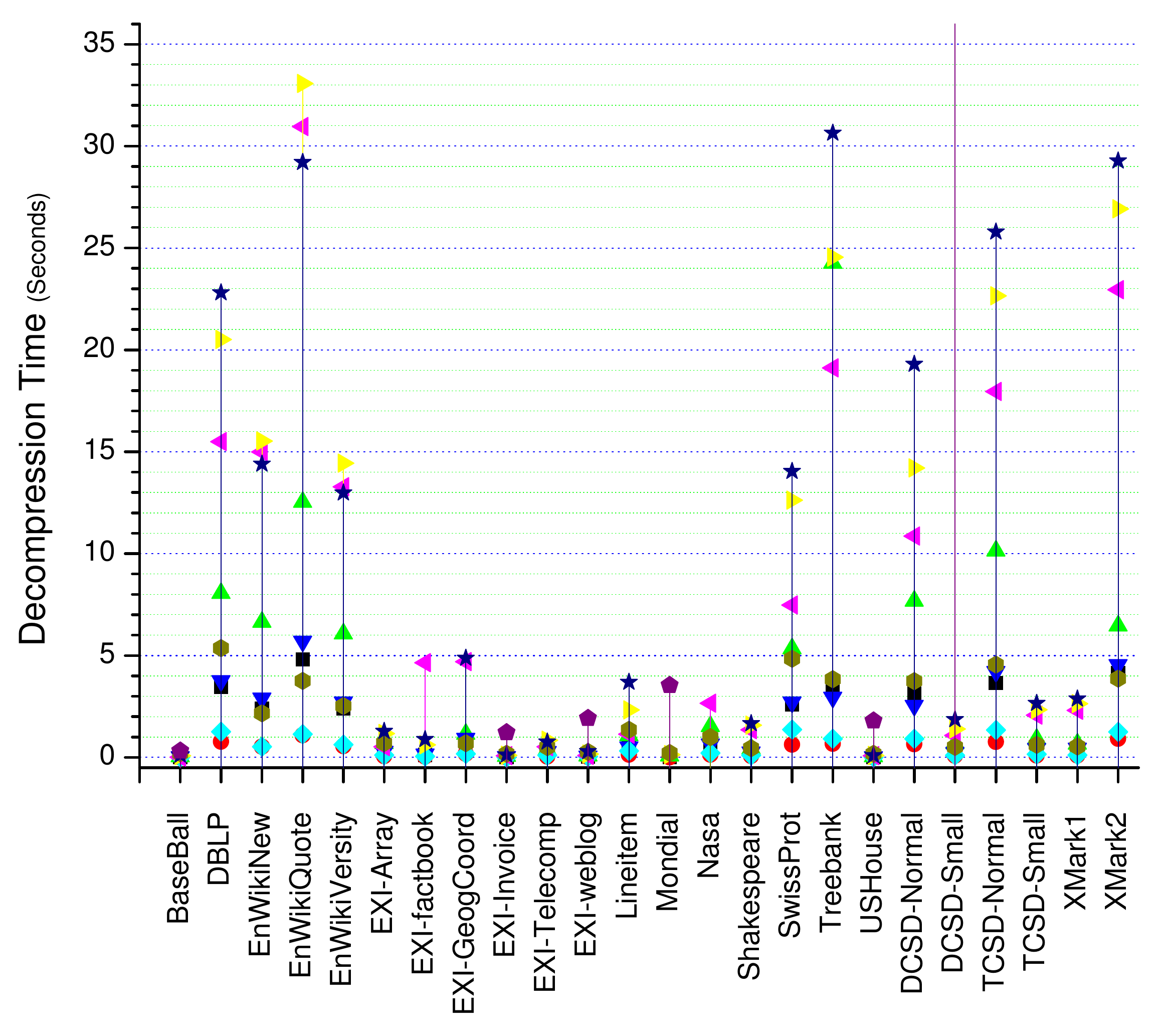}
\end{minipage}
\begin{minipage}[b]{0.35\textwidth}
\centering
\includegraphics[width=1\textwidth,height=3.3in]{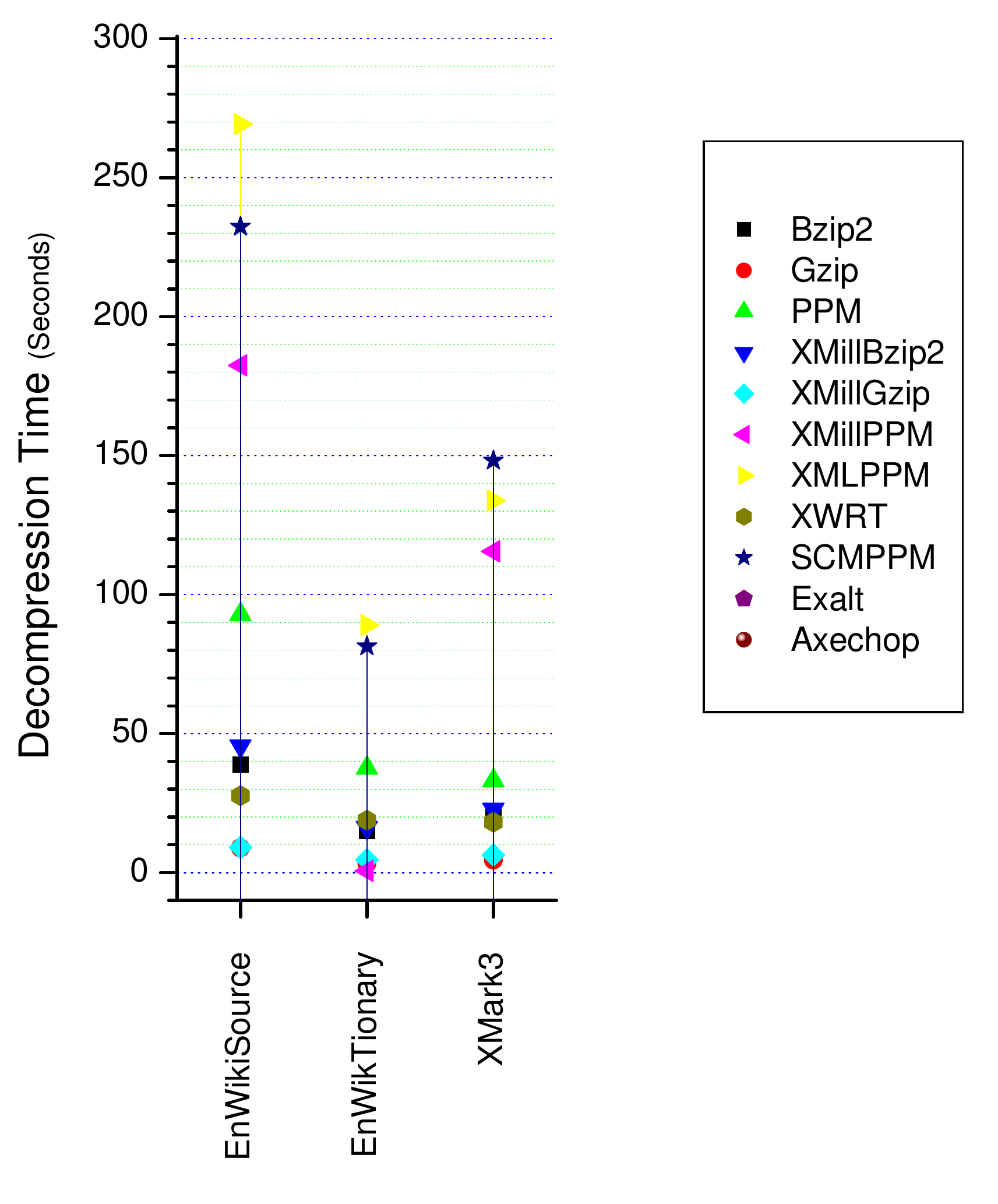}
\end{minipage}
\caption{Detailed decompression times on the high resources
setup.} \label{FIG-S-O-DCT-D}
\end{figure*}

\begin{figure*}
\begin{minipage} [b]{0.62\textwidth}
\centering
\includegraphics[width=1\textwidth,height=3.3in]{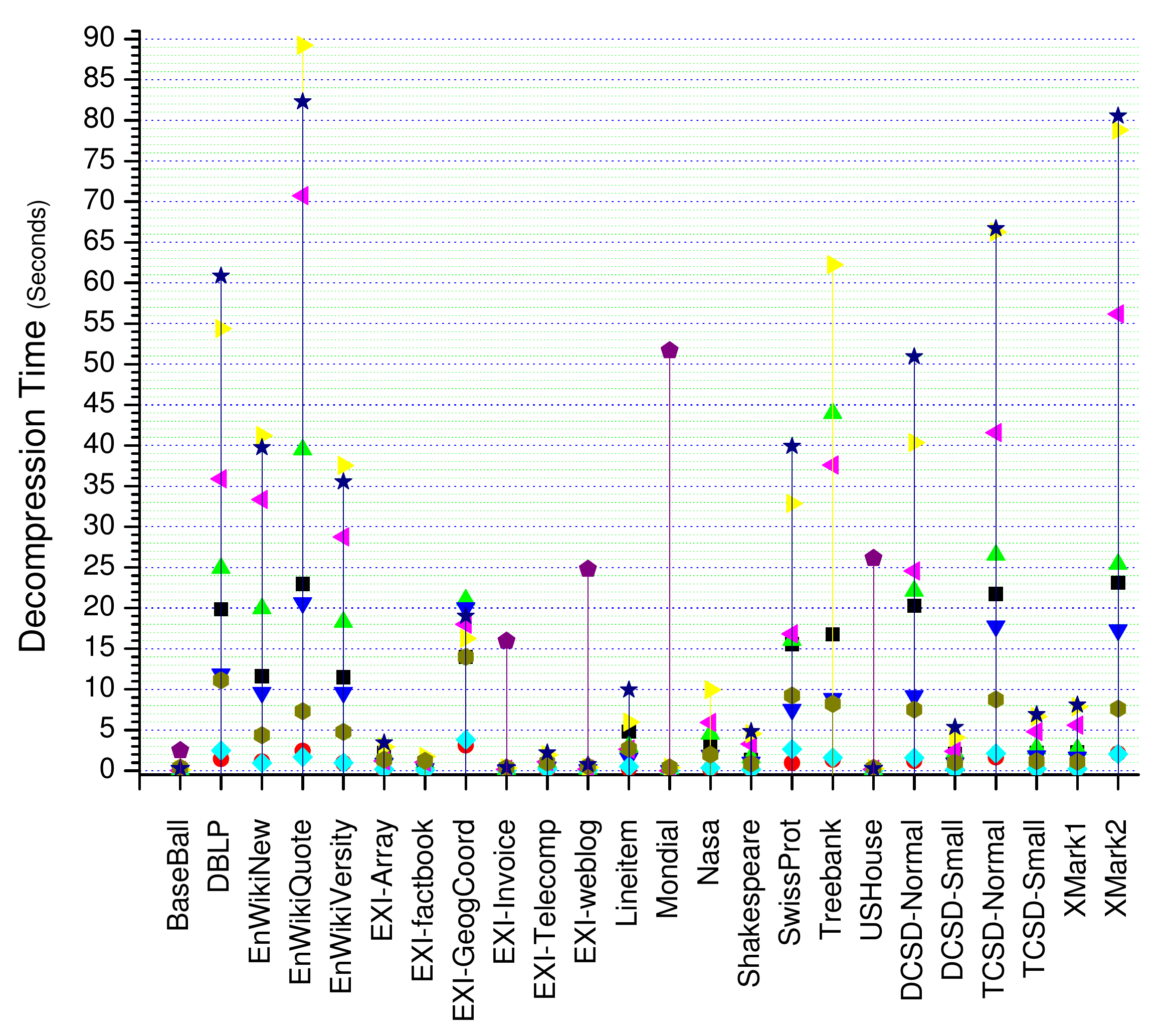}
\end{minipage}
\begin{minipage}[b]{0.35\textwidth}
\centering
\includegraphics[width=1\textwidth,height=3.3in]{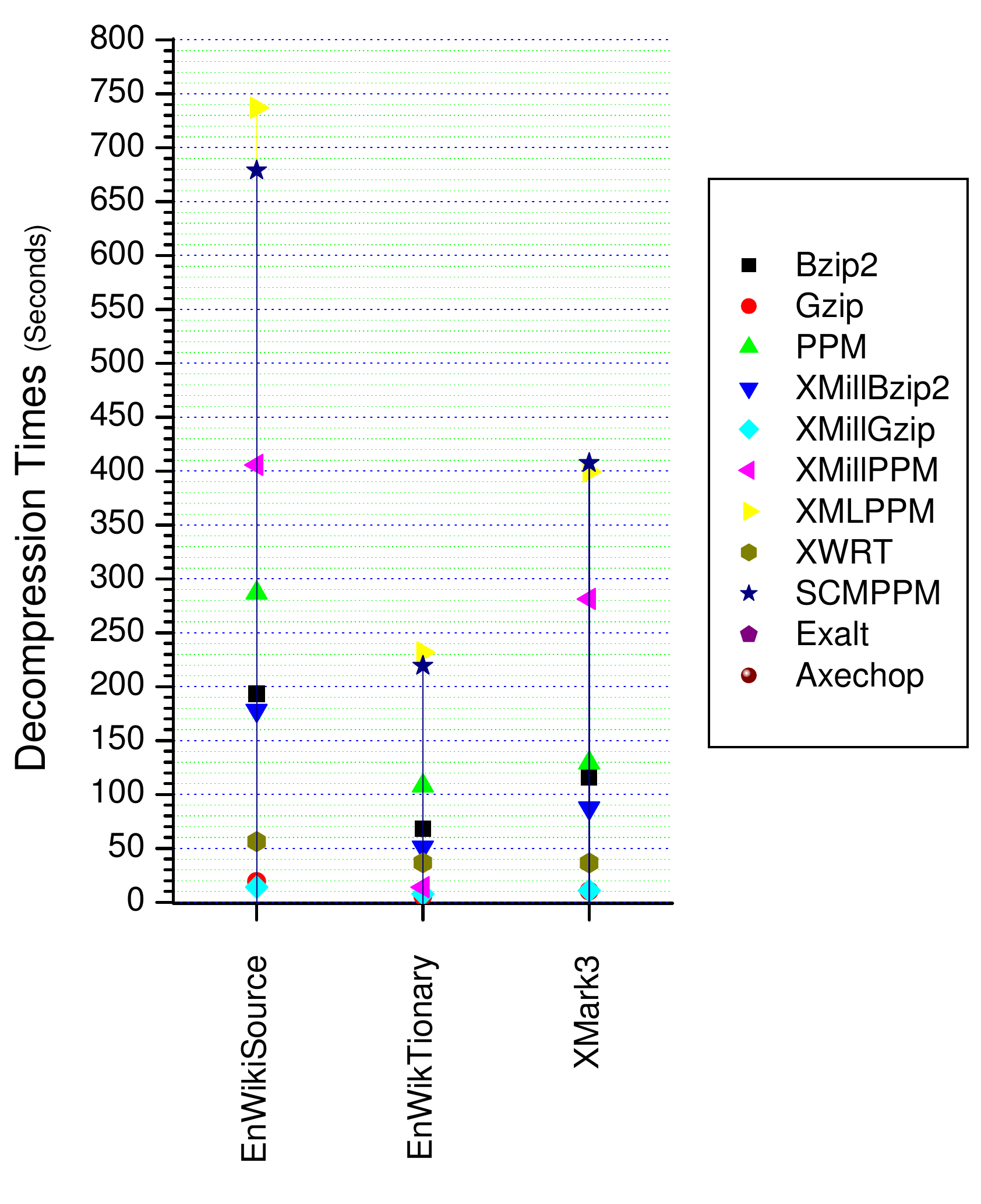}
\end{minipage}
\caption{Detailed decompression times on the limited resources
setup.} \label{FIG-C-O-DCT-D}
\end{figure*}

\begin{figure*}
\centering \subfigure[Structural documents.] {
    \label{FIG-S-S-CR-S}
    \includegraphics[width=0.45\textwidth,height=2.45in]{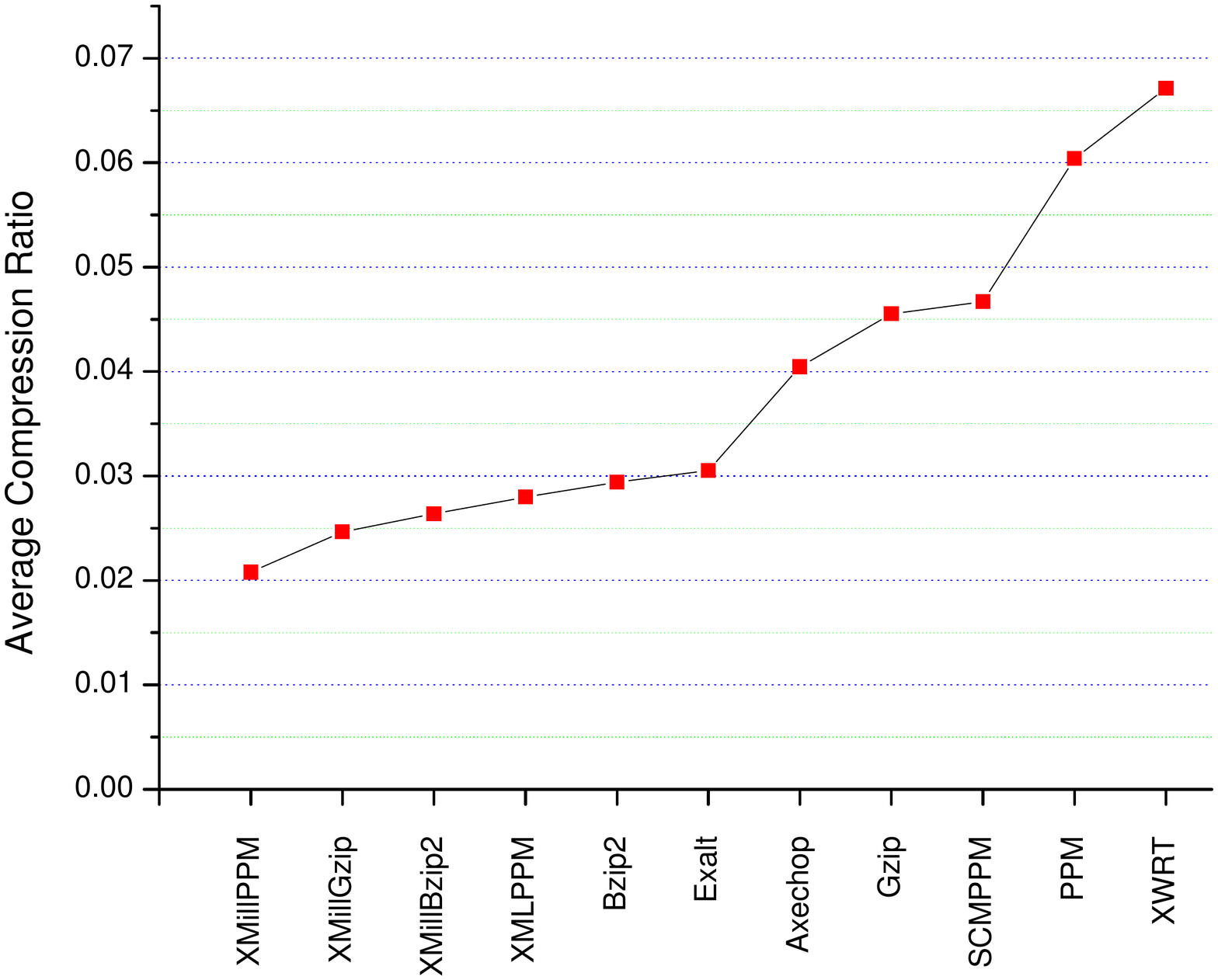}
}
\subfigure[Original documents.] {
    \label{FIG-S-O-CR-S}
    \includegraphics[width=0.45\textwidth,height=2.45in]{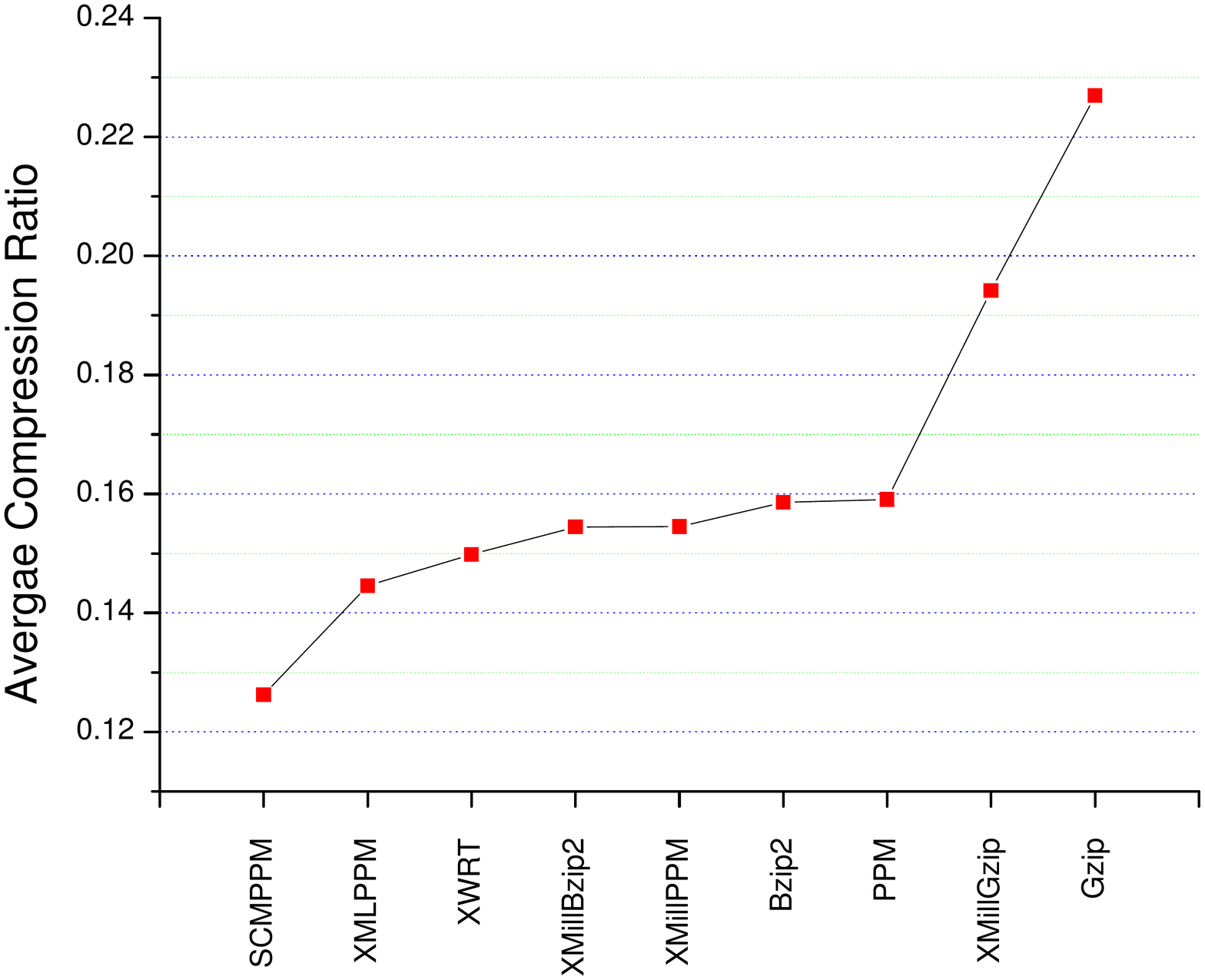}
} \caption{\footnotesize{Average compression ratios.}}
\label{FIG-S-O-CR-S}
\end{figure*}

\begin{figure*}
\centering \subfigure[Limited resources setup.] {
    \label{FIG-C-O-T-D}
    \includegraphics[width=0.45\textwidth,height=2.45in]{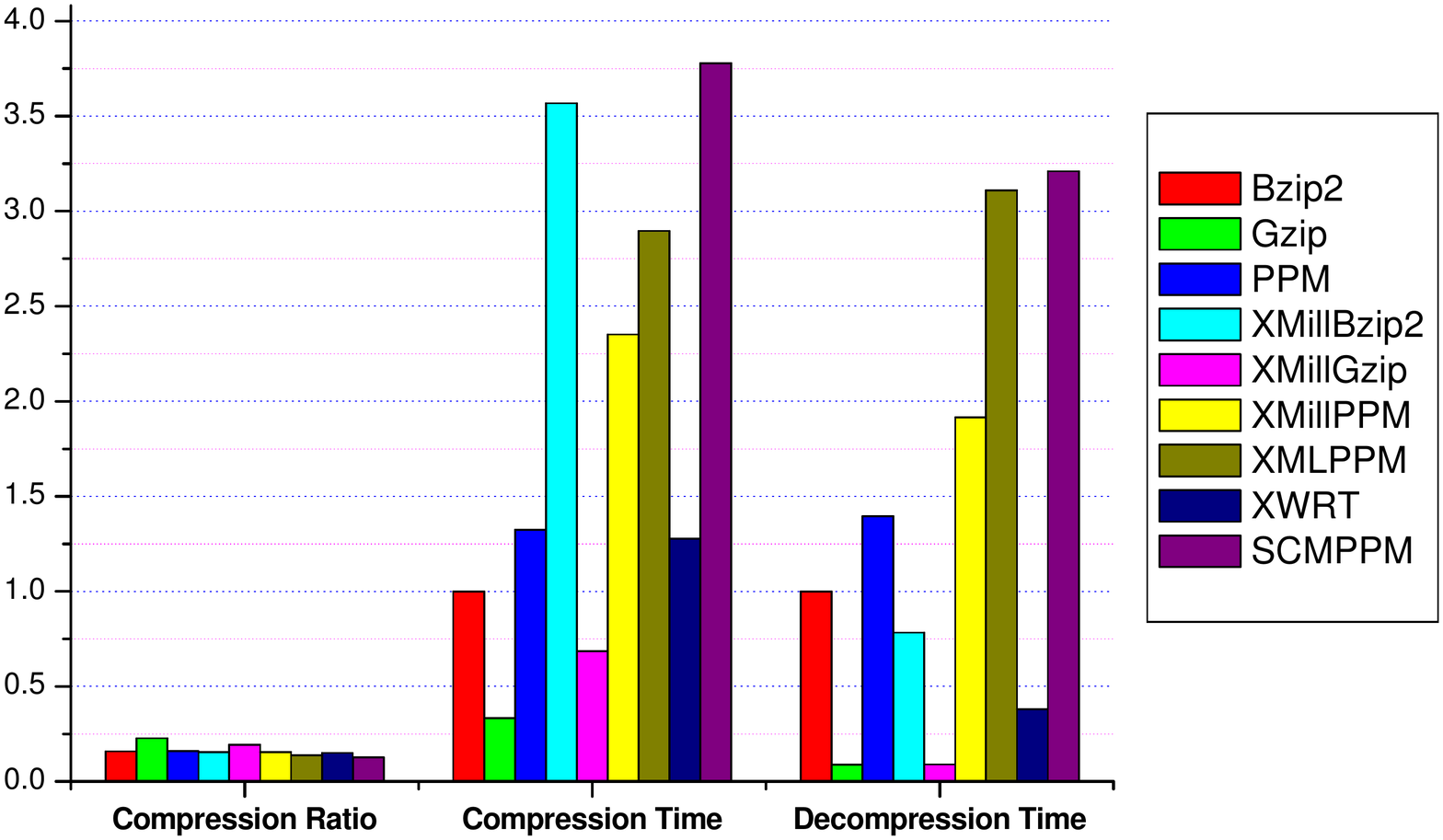}
}
\subfigure[High resources setup.] {
    \label{FIG-S-O-T-D}
    \includegraphics[width=0.45\textwidth,height=2.45in]{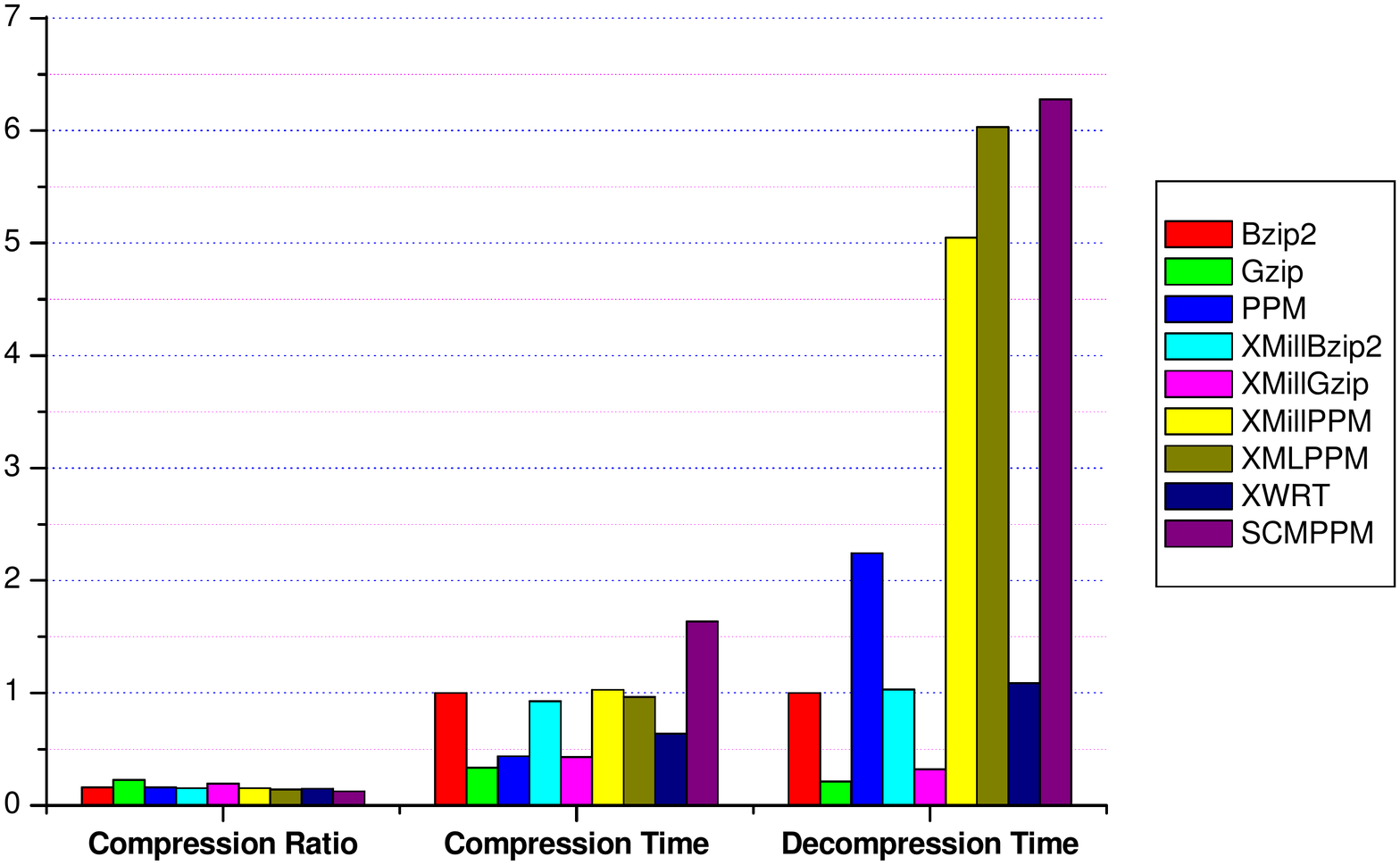}
} \caption{\footnotesize{Overall performance of compressing
\emph{original}  documents.}} \label{FIG-O-T-S}
\end{figure*}

In this section we report the results obtained by running our
exhaustive set of experiments. Figures \ref{FIG-P-O-CR-S} to
\ref{FIG-O-CR-S}  represents an important part of the results of
our experiments. Several remarks and guidelines can be observed
from the results of our exhaustive set of experiments. Some key
remarks are given as follows:\begin{itemize}

\item The results of Figure \ref{FIG-P-O-CR-S} and  Figure
\ref{FIG-P-S-O-T-D} show that the \emph{tuned} run of XWRT with
the highest level of compression ratio achieves the overall best
average compression ratio with very expensive cost terms of
compression and decompression times.

    \item Figure \ref{FIG-S-S-CR-S} shows that the three alternative back-ends of
\emph{XMill} compressor achieve the best compression ratio over
the structural documents. Figure \ref{FIG-S-S-CR-D} shows that
XMillPPM achieves the best compression ratio for all the datasets.
The irregular structural documents (Treebank, R1, R2, R3) are very
challenging to the set of the compressors. This explains why they
all had the worst compression ratios. \item Figures
\ref{FIG-S-O-CR-D} and \ref{FIG-S-O-CR-S} show that gzip-based
compressors (gzip, XMLGzip) have the worst compression ratios.
Excluding these two compressors, Figure \ref{FIG-S-O-CR-S} shows
that the differences on the average compression ratios between the
rest of compressors are very narrow. They are very close to each
other, the difference between the best and the worst average
compression ratios is less than 5\%. Among all compressors, SCMPPM
achieves the best average compression ratio. \item Figures
\ref{FIG-S-O-CT-D},\ref{FIG-C-O-CT-D},\ref{FIG-S-O-DCT-D},\ref{FIG-C-O-DCT-D}
show that the gzip-based compressors have the best performance in
terms of compression time and decompression time metrics on both
testing environments. The compression and decompression times of
the PPM-Based compression scheme (XMillPPM, XMLPPM, SCMPPM) are
much slower than the other compressors. Among all compressors,
SCMPPM has the longest compression and decompression times. \item
Figure \ref{FIG-O-T-S} illustrates the overall performance of XML
compressors on the high and limited resources setup where the
values of the \emph{performance metrics} are \emph{normalized}
respect to bzip2. The results of this figure illustrate the narrow
differences between the XML compressors in terms of their
compression ratios and the wide differences in terms of their
compression and decompression times.

\end{itemize}
\subsection{Ranking}
Obviously, it is a nice idea to use the results of our experiments
and our performance metrics to provide a global ranking of XML
compression tools. This is however an especially hard task. In
fact, the results of our experiments have not shown a \emph{clear
winner}. Hence, different ranking methods and different weights
for the factors could be used for this task. Deciding the weight
of each metric is mainly dependant on the scenarios and
requirements of the applications where these compression tools
could be used. In this paper we used three ranking functions which
give different weights for our performance metrics. These three
rankings function are defined as follows:
\begin{itemize}
    \item $WF1 = (1/3 * CR) + (1/3 * CT) + (1/3 * DCT)$.  \item
$WF2 =  (1/2 * CR) + (1/4 * CT) + (1/4 * DCT)$ \item $WF3 = (3/5 *
CR) + (1/5 * CT) + (1/5 * DCT)$  \end{itemize} where $CR$
represents the compression ratio metric, $CT$ represents the
compression time metric and $DCT$ represents the decompression
time metric. In these ranking functions we used increasing weights
for the compression ratio ($CR$) metric (33\%, 50\% and 60\%)
while $CT$ and $DCT$ were equally sharing the remaining weight
percentage for each function. Figure \ref{FIG-O-CR-S} shows that
\emph{gzip} and \emph{XMLGzip} are ranked as the best compressors
using the three ranking functions and on both of the testing
environments. In addition, Figure \ref{FIG-O-CR-S} illustrates
that none of the XML compression tools has shown a significant or
noticeable improvement with respect to the compression ratio
metric. The increasing assignment for the weight of $CR$ do not
change the order of the global ranking between the three ranking
functions.

\section{Conclusion}
\label{SecConclusion} 
\begin{figure*}
\centering \subfigure[Limited resources setup.] {
    \label{FIG-C-O-T-S}
    \includegraphics[width=0.45\textwidth,height=2.45in]{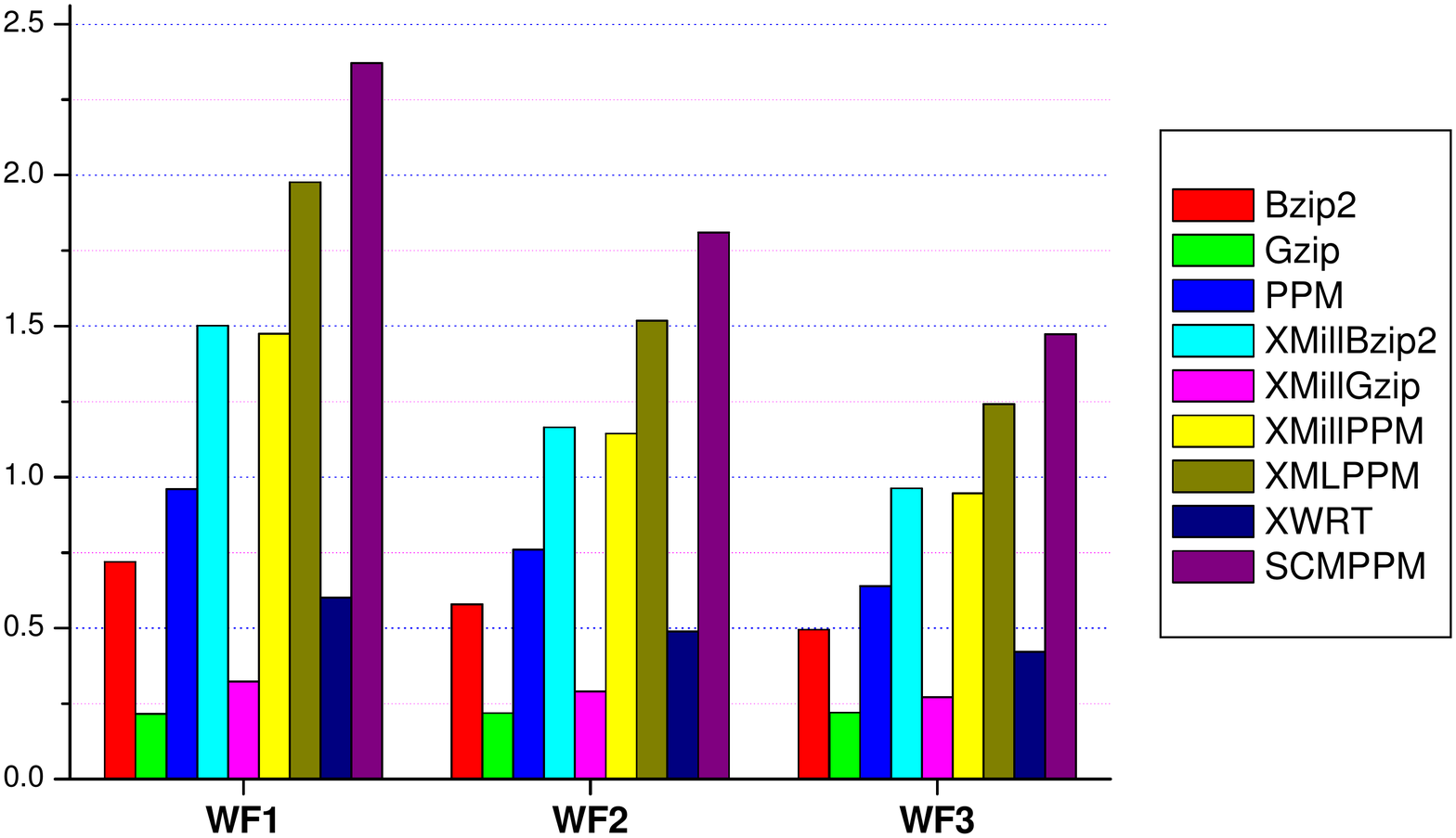}
}
\subfigure[High resources setup.] {
    \label{FIG-S-O-T-S}
    \includegraphics[width=0.45\textwidth,height=2.45in]{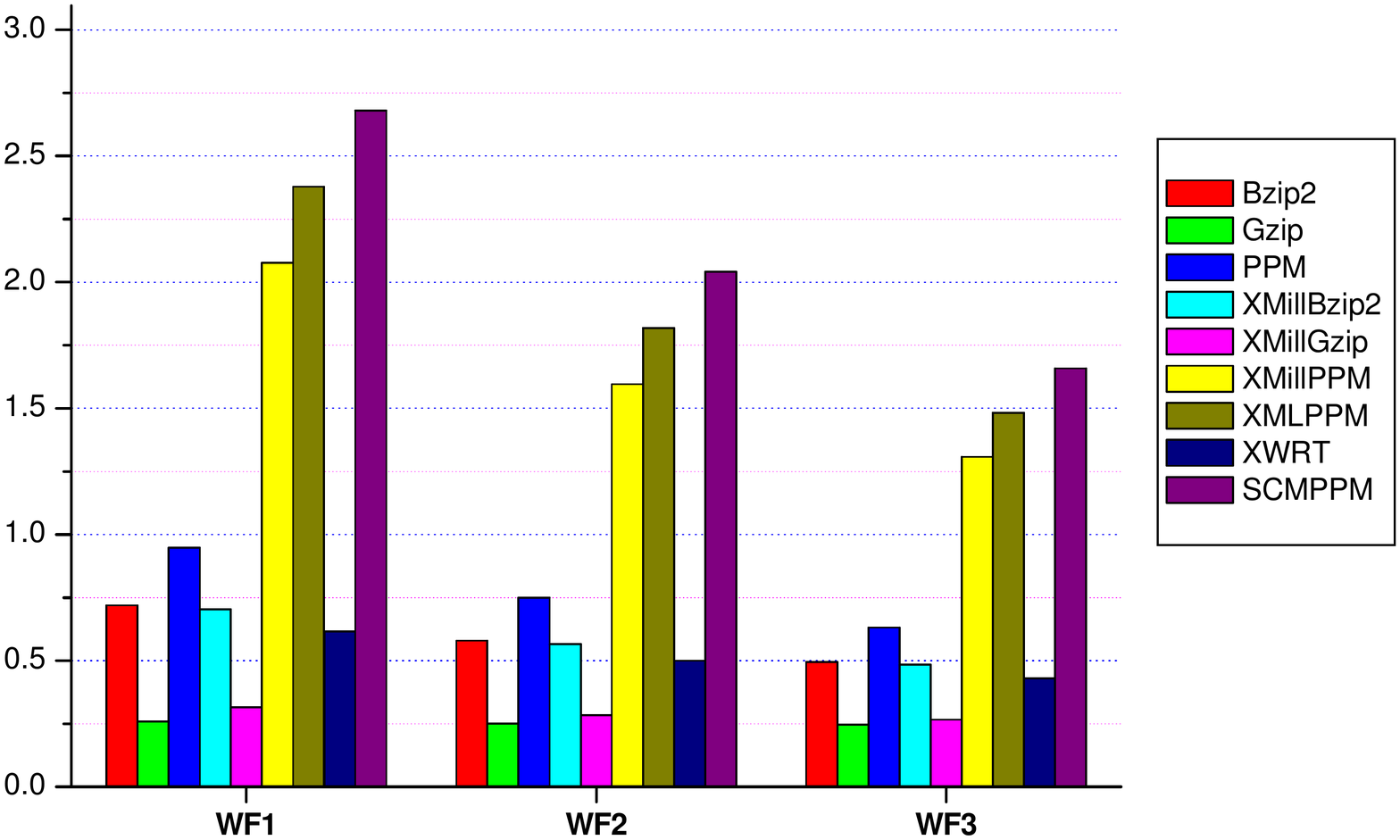}
} \caption{\footnotesize{Ranking Functions of compressing
\emph{original} documents.}} \label{FIG-O-CR-S}
\end{figure*}

We believe that this paper could be valuable
for both the developers of new XML compression tools and
interested users as well. For developers, they can use the results
of this paper to effectively decide on the points which can be
improved in order to make an effective contribution. For this
category of readers, we recommend tackling the area of developing
stable efficient \emph{queriable} XML compressors. Although there
has been a lot of literature  presented in this domain, our
experience from this study lead us to the result that we are still
missing efficient, scalable and stable implementations in this
domain. For users, this study could be helpful for making an
effective decision to select the suitable compressor for their
requirements. For example, for users with highest compression
ratio requirement, the results of Figure \ref{FIG-P-O-CR-S}
recommends the usage of either the PPM compressor with the highest
level of compression parameter (\emph{ppmd e -o16 document.xml})
or the XWRT compressor with the highest level of compression
parameter (\emph{xwrt -l14 document.xml})(if they have more than 1
GB RAM on their systems) while for the users with fastest
compression time and moderate compression ratio requirements, gzip
and XMillGzip are considered to be the best choice (Figure
\ref{FIG-O-CR-S}).

From the experience and the results of this experimental study, we
can draw the following conclusions and recommendations:
\begin{itemize}
    \item The primary innovation  in the XML compression mechanisms was
presented in the first implementation in this domain by XMill. It
introduced the idea of separating the structural part of the XML
document from the data part and then group the related data items
into homogenous containers that can be compressed separably. This
separation improves the further steps of compressing these
homogenous containers using the general purpose compressors or any
other compression mechanism because they can detect the redundant
data easily. Most of the following XML compressors have simulated
this idea in  different ways.

\item The dominant practice in most of the XML compressors is to
utilize the well-known structure of XML documents for applying a
pre-processing encoding step and then forwarding the results of
this step to general purpose compressors. Consequently,  the
compression ratio of most XML conscious compressor is  very
dependent and related on the general purpose compressors such as:
gzip, bzip2 or PPM. Figure \ref{FIG-S-O-CR-S} shows that none of
the  XML conscious compressors has achieved an outstanding
compression ratio over its back-end general purpose compressor.
The improvements are always not significant with 5\% being the
best of cases. This fact could explain why XML conscious
compressors are not widely used in practice.

\item The compression time and decompression time metrics play a
crucial role in the ranking of XML compressors.

\item The authors of the XML compression tools should provide more
attention to provide the source code of their implementations
available. Many tools presented in the literature - specially the
queriable ones - have no available source code which prevents the
possibility of ensuring the repeatability of the reported numbers.
It also hinders the possibility of performing fair and consistent
comparisons between the different approaches. For example in
\cite{XseqPaper}, the authors compared the results of their
implementation \emph{Xseq} with \emph{XBzip} using an inconsistent
way. They used the reported query evaluation time of \emph{XBzip}
in \cite{XBZIPPaper} to compare with their times although each of
the implementation is running on a different environment.

\item  There are no publicly available  solid implementations for
grammar-based XML compression techniques and queriable XML
compressors. These two areas provide many interesting avenues for
further research and development.

\end{itemize}

As a future work, we are planning to continue maintaining and
updating the web page of this study with further evaluations of
any new evolving XML compressors. In addition, we will enable the
visitor of our web page to perform their online experiments using
the set of the available compressors and their own XML documents.

\bibliographystyle{plain}
\bibliography{Biblio}

\small

\end{document}